\documentclass[onecolumn]{aastex6}
\usepackage[bottom=1cm]{geometry}
\usepackage{graphicx}
\usepackage{epsfig}
\begin{document}
\title{SMBH Seeds: Model Discrimination with High Energy Emission Based on Scaling Relation Evolution}
\author{Sagi Ben-Ami}
\affiliation{Harvard-Smithsonian Center for Astrophysics, 60 Garden Street, Cambridge, MA 02138, USA}
\author{Alexey Vikhlinin}
\affiliation{Harvard-Smithsonian Center for Astrophysics, 60 Garden Street, Cambridge, MA 02138, USA}
\author{Abraham Loeb}
\affiliation{Institute for Theory and Computation, Harvard University, 60 Garden Street, Cambridge, MA, 02138, USA}

\begin{abstract}
We explore the expected X-ray (0.5-2keV) signatures from super massive black holes (SMBHs) at high redshifts ($z\sim5-12$) assuming various models for their seeding mechanism and evolution. The seeding models are approximated through deviations from the M$_{BH}-\sigma$ relation observed in the local universe. We use results from N-body simulations of the large-scale structure to estimate the density of observable SMBHs.
We focus on two families of seeding models: (\textit{i}) light seed BHs from remnants of Pop-III stars; and (\textit{ii}) heavy seeds from the direct collapse of gas clouds. We investigate several models for the accretion history, such as sub-Eddington accretion, slim disk models allowing mild super-Eddington accretion and torque-limited growth models. We consider observations with two instruments: (\textit{i}) the Chandra X-ray observatory, and (\textit{ii}) the proposed Lynx. We find that all the simulated models are in agreement with the current results from Chandra Deep Field South (CDFS) - \textit{i.e.,} consistent with zero to a few observed SMBHs in the field of view. In deep Lynx exposures, the number of observed objects is expected to become statistically significant. We demonstrate the capability to limit the phase space of plausible scenarios of the birth and evolution of SMBHs by performing deep observations at a flux limit of $1\times10^{-19}\mathrm{erg\,cm^{-2}\,s^{-1}}$. Finally, we estimate the expected contribution from each model to the unresolved cosmic X-ray background (CXRB), and show that our models are in agreement with current limits on the CXRB and the expected contribution from unresolved quasars. We find that an analysis of CXRB contributions down to the Lynx confusion limit yields valuable information that can help identify the correct scenario for the birth and evolution of SMBHs.
\end{abstract}
\section{Introduction}
Data accumulated over the past three decades suggest that most of the galaxies host a quiescent supermassive black hole (SMBH) at the center \citep{2005SSRv..116..523F}. The mass of the observed SMBHs is well correlated with properties of the host spheroid such as its stellar component velocity dispersion and luminosity \citep{1995ARA&A..33..581K,2009ApJ...698..198G}. The observed correlations point to co-evolution of SMBHs and their host galaxies, most likely through feedback effects from the SMBHs that result in regulation of the cold gas supply in galactic nuclei, \citep[and reference therein]{Volonteri2010}. One of the first scenarios suggested for the birth of SMBH seeds was the collapse of  population III (Pop-III) stars. These stars are assumed to have a top-heavy stellar initial mass function (IMF) and very low metallicity \citep{Bromm1999,Gao2007}, and so would have short lifetimes of a few Myrs while being able to retain most of their mass throughout their evolution. Eventually these stars will collapse directly into $\sim100$M$_{\odot}$ black holes (BHs) due to loss of pressure as they consume their nuclear fuel \citep{Heger2003}. As the host galaxies evolve and merge, the BHs grow through accretion and mergers, as new supply of cold gas is flowing into their host central region.

The detection of quasars at high redshifts has posed challenges to the Pop-III scenario. In order to grow a $100$M$_{\odot}$ BH to a $10^9$M$_{\odot}$ SMBH in less than $\sim1\,$Gyr (the current record holder, ULAS J1120+0641 at a redshift of $z\sim7.1$, merely $770\,$Myr after the big bang, has an estimated mass of $2^{+1.5}_{-0.7}\times10^9$M$_{\odot}$, \cite{2011Natur.474..616M}), we need to assume a continuous accretion at the Eddington limit during most of its lifetime \citep[\textit{e.g.,}][]{Haiman2001}. This is an unlikely premise due to expected feedback effects from the accreting  massive black hole (MBH). A possible solution is to assume the black hole undergoes short yet strong  growth episodes during which its accretion rate is well in excess of its Eddington accretion rate  \citep[\textit{e.g.,}][]{Wyithe2012,Madau2014,Alexander2014}. This solution raises further difficulties as the capability of a BH to sustain super-Eddington accretion without dispersing the cold gas streams that feeds it is debatable. Another problem encountered by the Pop-III scenario results from recent simulations for the formation of the first generation of stars \citep{Clark2008,Turk2009}. These suggest that the IMF is not as top heavy as initially expected, and it is no longer clear whether enough objects will end their lives in a direct collapse to a BH to account for the density of SMBHs in the local universe \citep{Volonteri2010}.

The difficulties raised by the Pop-III scenario (hereafter 'light seed models') in explaining the presence of quasars at high redshifts have prompted searches for alternative channels for the birth of SMBH seeds. Models such as the direct collapse of a gas cloud to a MBH \citep[\textit{e.g.,}][]{Haehnelt1993, Loeb1994, Eisenstein1995,Bromm2003}, or the generation of a giant proto-star that further collapses to a MBH, are accepted today as viable solutions. These scenarios advocate that SMBHs originate from heavy seeds, with masses of $10^5-10^6$M$_{\odot}$; hereafter we label this family of models 'heavy seed models'. Since the growth time of a SMBH is inverse proportional to its seed mass \citep[\textit{e.g.,}][]{Haiman2001}, this family of models alleviates many of the problems encountered by the Pop-III scenario.

Despite the growing interest in the field, there is still no clear evidence for the correct scenario for the birth of SMBH seeds. A key problem encountered when looking for evidence in the local universe is that signatures of the seeding mechanism can be washed out during  the SMBH - host galaxy co-evolution \citep{Natarajan2011}. Moreover, with current capabilities one observes only massive, high luminosity objects at high redshift, \textit{i.e.,} those which have already undergone major growth past the seed stage \citep{Merloni2016}. Many searches for the signatures of intermediate MBHs at high redshift have been performed so far, with a significant number of them in the X-ray band \citep[\textit{e.g.,}][]{Vito2013, Weigel2015}. A fundamental advantage of X-ray observations is that the emitted photons are in general energetic enough to escape dust and gas clouds surrounding the MBH, and therefore can give us a full census of the flux distribution, related to the MBH accretion rate and mass. While flux limits from current X-ray observatories such as Chandra and XMM-Newton can in general detect a MBH with a mass of $\sim10^6$M$_{\odot}$ up to a redshift of $z\sim7$ (see section 2), no such objects have been robustly detected even in the deepest surveys \citep[\textit{e.g.,}][]{Treister2013}.

In this paper we explore the X-ray flux distribution from MBHs at high redshifts as they emerge from the seed state and co-evolve with the host galaxy. We investigate whether current or future experiments can shed light on the mechanism responsible for the birth of SMBH seeds at high redshift. By using physically motivated models, we estimate the X-ray flux distribution from MBHs at high redshift and determine whether different seeding and growth models yield different signatures during the initial stages of MBH-host co-evolution when observed in the X-ray band. We focus on two observational programs: The Chandra Deep Field South survey \citep[CDFS;][]{Xue2011,Luo2017}, and a similar observing program with the proposed Lynx.  Due to its excellent spatial resolution (half power diameter, HPD, of 0.5'') Chandra is currently the best instrument for detecting low flux objects at the high redshifts we consider. CDFS has a point source detection limit of $9.1\times10^{-18}\mathrm{erg\,cm^{-2}\,s^{-1}}$ at energies of $0.5-2\,$keV in a $5'\times5'$ field of view (FoV). Lynx is a future X-ray observatory aiming at a spatial resolution similar to that of Chandra with an effective area orders of magnitude greater. Lynx \citep{Lynx} is designed to reach a flux limit of $\sim10^{-19}\mathrm{erg\,cm^{-2}\,s^{-1}}$ over $400\,\mathrm{arcmin}^2$ in its deepest surveys. 

Our paper is organized as follows. In section II we describe the basic assumptions in our models, discuss the reasoning behind them, and present the methodology used to estimate the expected observed X-ray flux distributions for the experiments under study. Section III gives an analysis of the results with an emphasis on model discrimination given the derived X-ray flux distributions. In section IV we investigate models allowing super-Eddington accretion. A summary and conclusions are given in section V. Throughout the paper, we assume standard $\Lambda$CDM cosmology with $h=0.7,\ \Omega_m=0.27$, and $\sigma_8=0.82$, in agreement with current Planck measurements \citep{Planck2016} and the parameters used in several of the leading numerical simulations \cite[\textit{e.g.,}][]{Klypin2011}.
\newpage
\section{Methodology}
Our goal is to investigate whether the detected high-energy emission with current (Chandra) and future (the proposed Lynx) telescopes allows us to discriminate between the various scenarios for the birth and growth of SMBHs. We first determine the minimum mass of a MBH that can be detected with a given experiment as a function of the luminosity scaling parameter $\lambda$, under favorable conditions (\textit{e.g.,} no circum-nuclear,  ISM, and IGM absorption). Guided by merger tree simulations for the growth of SMBHs, we estimate the evolution of the M$_{BH}-\sigma$ relation at high redshifts, which allows us to relate MBHs to their host dark matter (DM) halos. Using cosmological simulations to estimate the number density of DM halos, we derive the expected number of observable MBHs at a given redshift for each seeding model and $\lambda$. The derived numbers provide input parameters to our Monte Carlo (MC) simulation that estimates the expected X-ray flux distributions from MBHs. The number of observed MBHs and their X-ray flux distributions are the observables in the 2-D phase space we study, a phase space in which one axis is a parameter related to the birth mechanism and the other axis is a luminosity scaling parameter related to the MBH accretion rate.\\
In this work we focus on the post seeding stage and the initial co-evolution of the MBH with its host galaxy. Therefore we do not attempt to model SMBH seeds formation, but assume all models are encoded via deviations from the M$_{BH}-\sigma$ relation observed in the local universe (\textit{i.e.,} a modified model and redshift dependent M$_{BH}-\sigma$ relation). The co-evolution stage is expected to be significantly longer than the seeding stage \citep[$\gtrsim1\,$Gyr vs.  $\sim0.1\,$Gyr; see][]{Volonteri2010,Agarwal2013,Natarajan2017}, and might offer a significant statistic in our effort to study the discriminating power of X-ray flux distribution for a wide range of models. In some cases, X-ray emission at the seed stage is expected to be detected \citep{Pacucci2015}. While this can give additional discrimination power, we do not consider it in this work. A detailed description for each step in our methodology follows.
\subsection{Black Hole Mass and Flux Limits }
\begin{figure}[]
\centering
{\includegraphics[width=0.475\columnwidth]{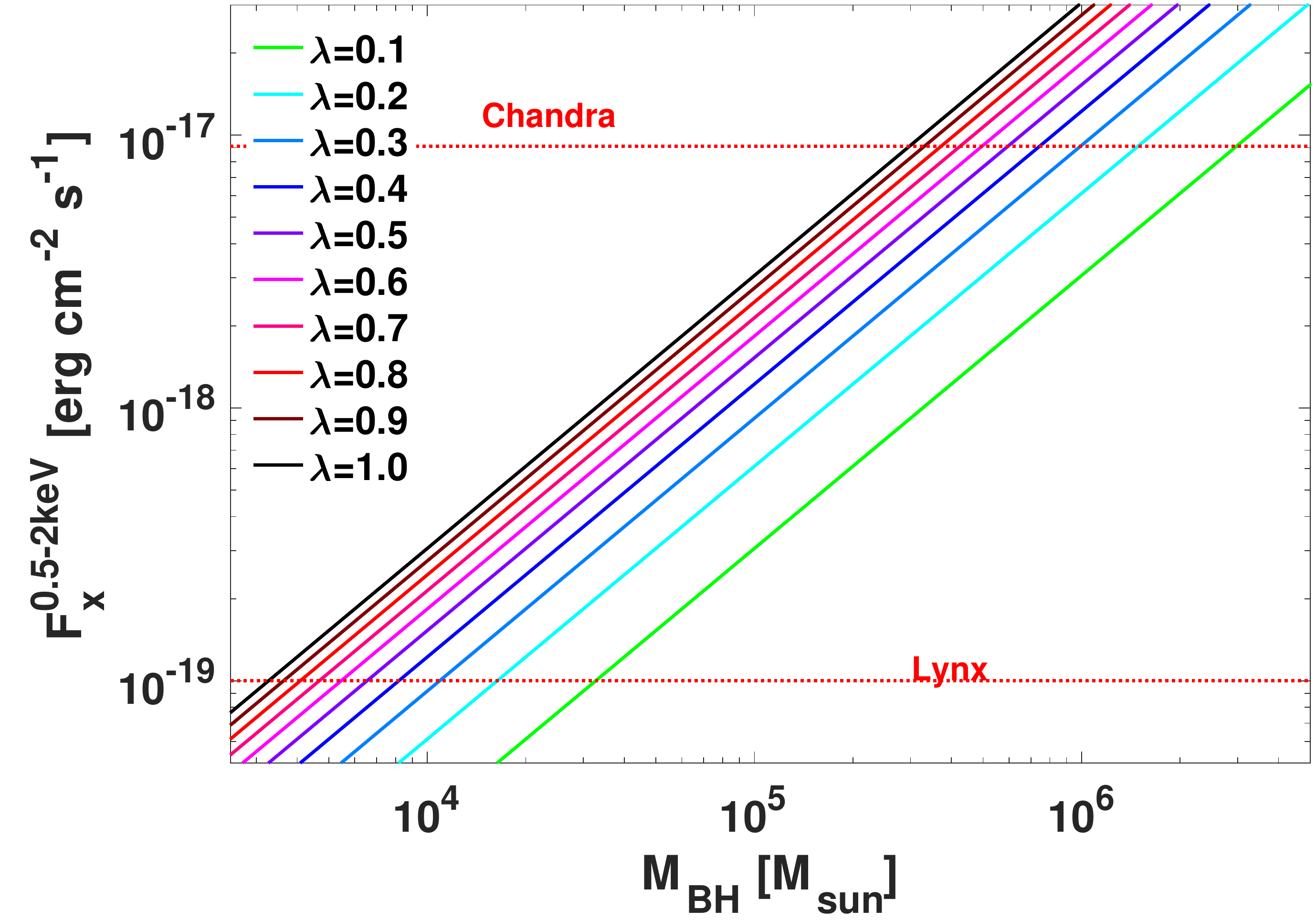}} 
{\includegraphics[width=0.475\columnwidth]{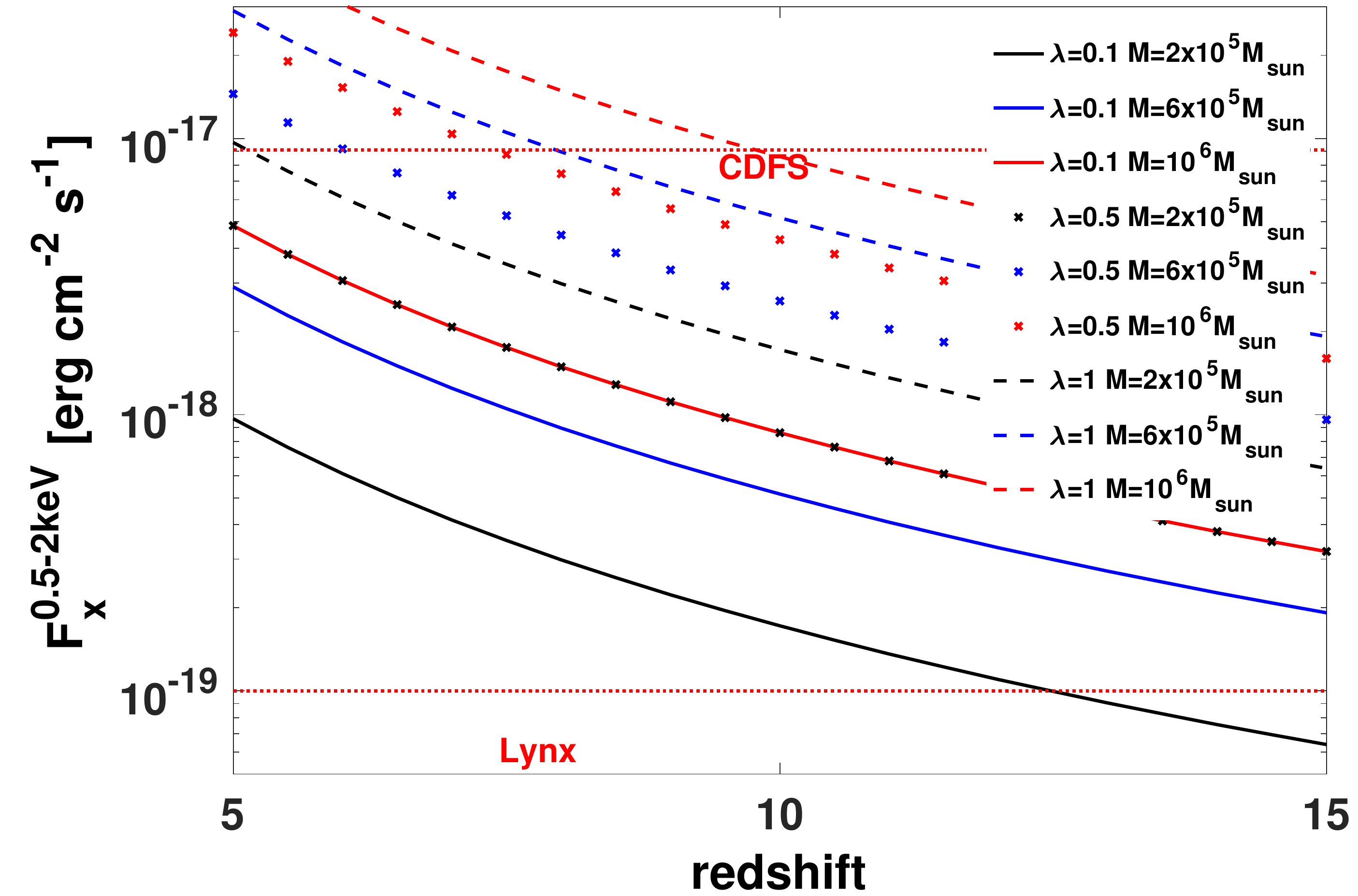}} \\
\caption{X-ray flux dependencies. Left: Observed X-ray flux between $0.5-2\,$KeV for MBHs at $z=6$, assuming sub-Eddington accretion characterized by the scale parameter $\lambda$. For Chandra/Lynx, we are able to detect MBHs of M$_{BH}>3\times10^5/\sim1\times10^4$M$_{\odot}$ ($\lambda=1.0$) to M$_{BH}>3\times10^6/\sim1\times10^5$ M$_{\odot}$ ($\lambda=0.1$). Right: X-ray flux as a function of redshift for BH masses of $2\cdot10^5-10^6\mathrm{M}_{\odot}$ in the sub-Eddington regime.}
\label{fig:FxToMass}
\end{figure}

The minimum mass of an observable MBH will depend on the object's emitted X-ray luminosity $L_{x}$, and its luminosity distance $D_L$ (see section 2.4 for other factors that might affect a MBH observed flux). Initially, we focus on sub-Eddington luminosities, so that the BH bolometric luminosity is given by $L_{BH}^{bol}=\lambda L_{edd}$. $L_{edd}$ is the BH Eddington luminosity, $L_{edd}=3.2\times10^4\left(\mathrm{M}_{BH}/\mathrm{M}_{\odot}\right)L_{\odot}$ \citep{Rybicki1979}, and $0<\lambda\leq1$ is a scaling parameter. We focus on rest frame energies of $3-26\,$keV, as photons of such energies emitted at $z\sim5-12$ \citep[initial co-evolution of host and SMBH seed; ][]{Volonteri2010,Natarajan2017} will be redshifted to energies of $0.5-2\,$keV, the energy band with peak response in the experiments we investigate. We make a conservative assumption regarding the flux fraction emitted at energies above $2\,$keV $\epsilon_{2keV}=0.1$ \citep[\textit{e.g.,}][]{Hopkins2007}, and so the X-ray flux observed between $0.5-2\,$keV is given by:
\begin{equation}
F_x=\epsilon_{2keV}\frac{\lambda L_{edd}}{4\pi D_L^2}\times k_{corr}^z=3.2\times10^4\frac{\epsilon_{2keV}\lambda k_{corr}^z}{4\pi D_L^2}\left(\frac{\mathrm{M}_{BH}}{\mathrm{M}_{\odot}}\right)L_{\odot}
\end{equation}
The K-correction $k_{corr}^z$ depends on the spectral energy distribution of the emitted object, which we assume behaves like a power-low with an index of $1.8$ so that $f(E)\propto E^{-1.8}$ (\textit{e.g.,} Ishibashi \& Courvoisier 2010, Mateos et al. 2010).
Figure \ref{fig:FxToMass} shows the X-ray flux dependency on mass (left panel) and on redshift (right panel). We plot the detection limits of the CDFS survey \citep[$9.1\times10^{-18}\mathrm{erg\,cm^{-2}\,s^{-1}}$;][]{Xue2011} and the expected Lynx deep field detection limit  \citep[$1\times10^{-19}\mathrm{erg\,cm^{-2}\,s^{-1}}$;][]{Lynx} for comparison. Only MBHs with masses above $\simeq10^6\mathrm{M}_{\odot}$, luminosities approaching $L_{edd}$, and closer than redshift $z\sim7$ can be detected with Chandra. On the other hand, Lynx allows us to detect BHs with masses as low as $\sim10^4$M$_{\odot}$ at the same redshifts, providing access to the seed popoulation.

\subsection{A redshift dependent M$_{BH}-\sigma$ relation}
\begin{figure}[]
\centering
{\includegraphics[width=0.475\columnwidth]{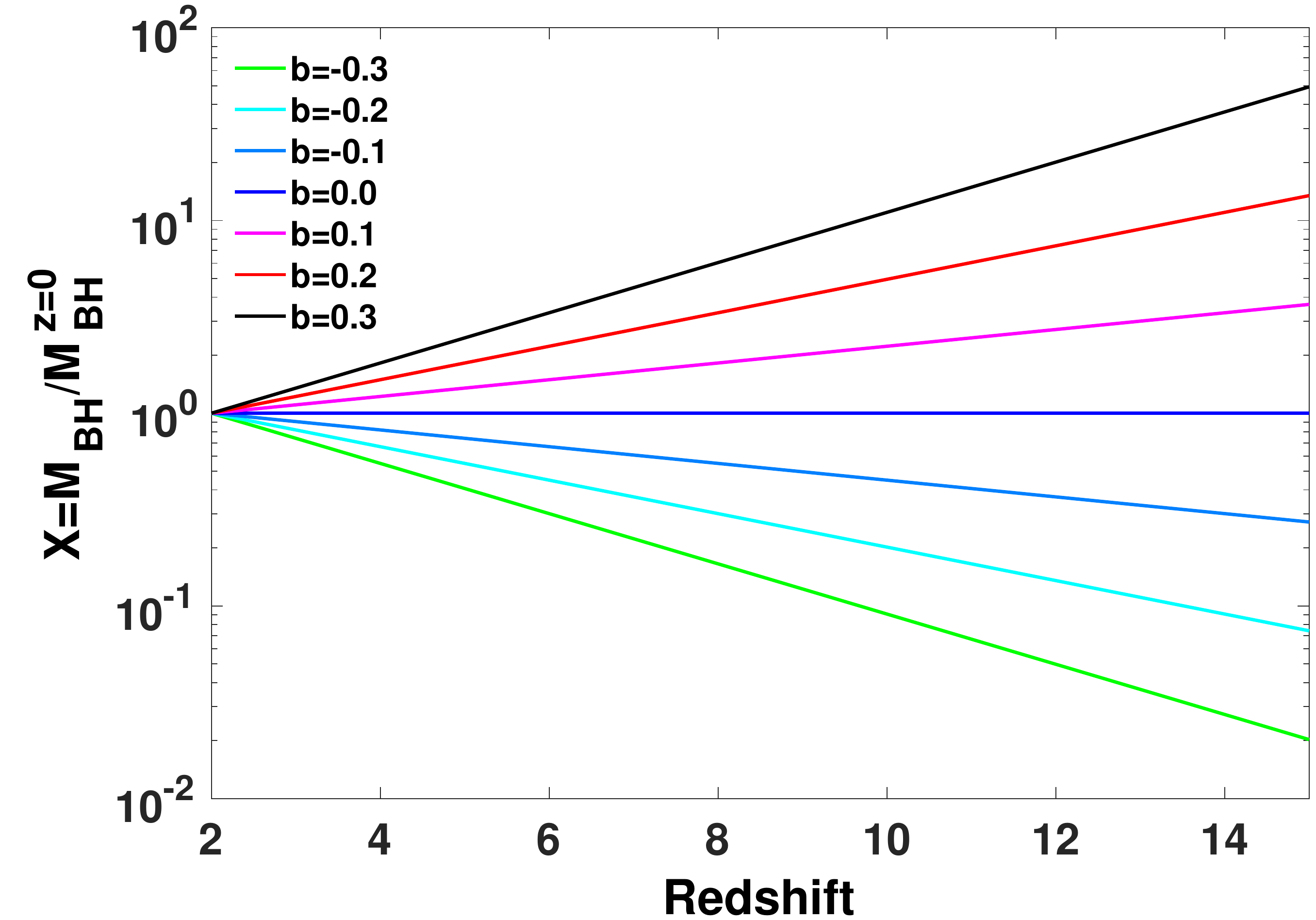}} 
{\includegraphics[width=0.475\columnwidth]{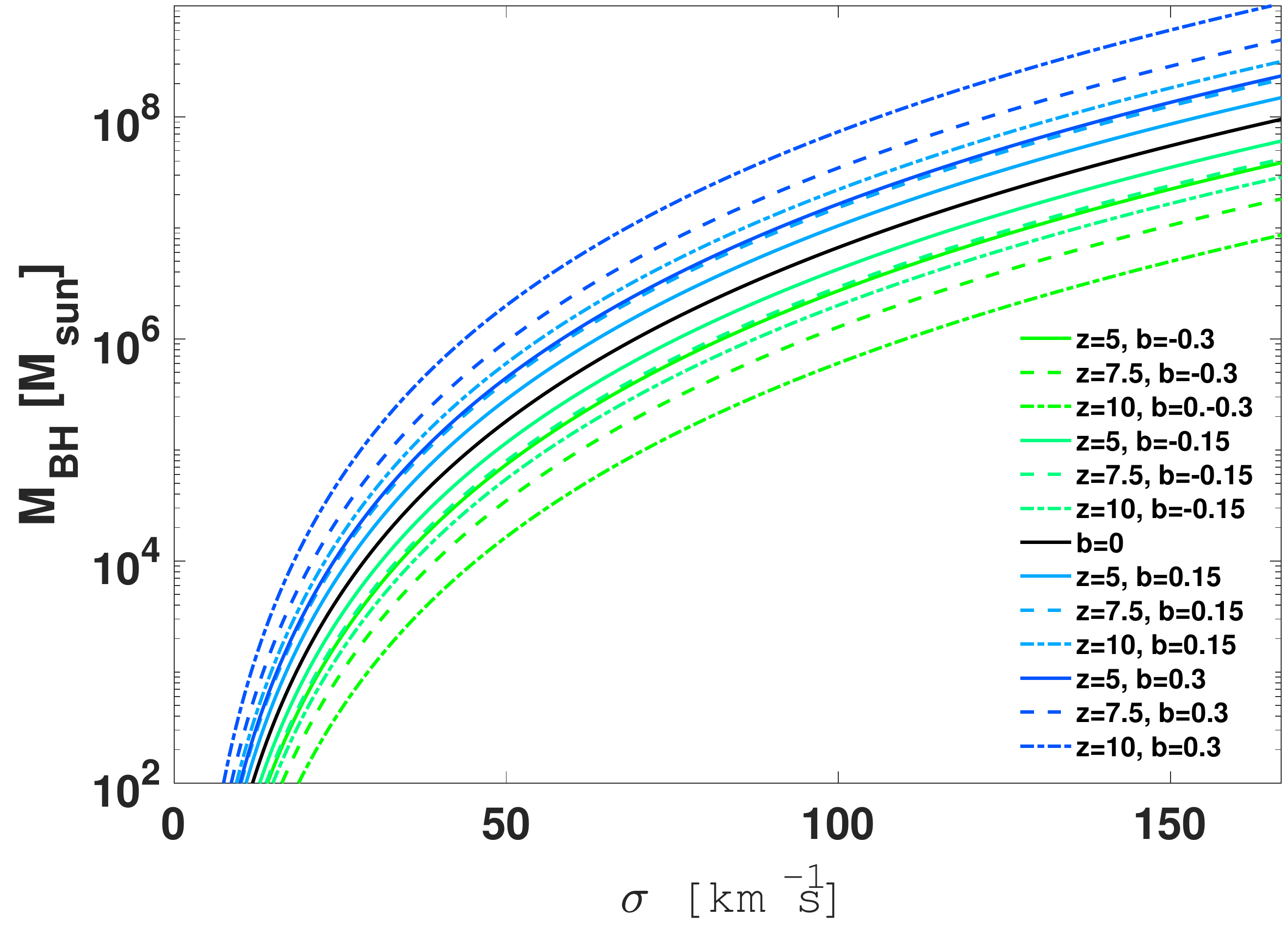}} \\
\caption{Modified M$_{BH}-\sigma$ relations. Left :Mass ratio parameter \textit{X}$=\mathrm{M}_{BH}^z/\mathrm{M}_{BH}^{z=0}$. We assume \textit{X}$=e^{b(z-2)}$. \textit{X}$>1$ indicates SMBH seeds originating from heavy seed models and are converging into the local M$-\sigma$ relation from above, while \textit{X}$<1$ indicates light seed models converging to the local M$-\sigma$ relation from below \citep[see text and][]{Volonteri2009}. Right: The modified M$_{BH}-\sigma$ relation for various models at various redshifts. A model with a positive (negative) value of the parameter $b$ implies that the MBH residing in the center of a DM halo with a specific value of $v_{circ}^{max}$ is heavier (lighter) than the MBH found at lower redshift in a DM halo of similar mass.}
\label{fig:MV_Relation}
\end{figure}

Empirical correlations between SMBH masses, and various properties of their host galaxies have been established over the past two decades. Power-law fits to these
correlations provide efficient means to estimate black hole masses in a statistically significant sample \citep{Beifiori2012,McConnell2013}. Several authors have argued the empirical scaling relations are rooted in a more fundamental relation between M$_{BH}$ and properties of the host galaxy dark matter halo, such as the halo virial mass (M$^{vir}_{DM}$) or $v_{circ}^{max}=\sqrt{\frac{GM(<r)}{r}}|_{max}$, the maximum circular velocity achieved beyond the bulge \citep[\textit{e.g.,}][]{Ferrarese2002,Saxton2014,Larkin2016}.  In this work we use $v_{circ}^{max}$ as provided by numerical simulations that trace DM halo evolution from primordial density fluctuations in combination with the correlation found between M$_{BH}$ and the stellar velocity dispersion of the bulge $\sigma$. We relate the mass of the black hole to $v_{circ}^{max}$ of the host, by assuming that the M$_{BH} - \sigma$ relation can be cast to an M$_{BH} - v_{circ}^{max}$ relation through a simple numerical factor $\eta=v_{circ}^{max}/\sigma$ that takes values of $2-4$ \citep{Padmanabhan2004,Dutton2004,Kravtsov2009}. We use the correlation found by \cite{McConnell2013} for late-type galaxies, $\log_{10}($M$_{BH})=8.07+5.06\log_{10}(\sigma/200\,\mathrm{km\,s^{-1}})$, as these are the majority of galaxies we expect at the observed redshifts.\\
The M$_{BH}-\sigma$ relation has been demonstrated across several orders of magnitude in SMBH masses in the local universe (\textit{i.e.,} $z\sim0$). Since we investigate MBHs after the seeding stage, and during initial co-evolution with the host, we assume a modified M$_{BH}-\sigma$ relation has already been established. Current observations, as well as theoretical modeling are inconclusive regarding the evolution of the M$_{BH}-\sigma$ relation at high redshifts. While several studies argue for a steepening of the slope in the M$_{BH}-\sigma$ relation \citep[\textit{e.g.,}][]{Treu2007,Targett2012}, other argue for no evolution with redshift \citep[\textit{e.g.,}][]{Jahnke2009, Schramm2013}, and even flattening of the slope at high redshifts \citep{Shapiro2009}. Any deviation from the local M$_{BH}-\sigma$ relation will encode physics of the seeding mechanism, and we therefore relate the M$_{BH}-\sigma$ relation at high redshift to the seeding mechanism and the MBH accretion history. 
Merger driven BH growth simulations, which track the growth of MBHs and their host galaxies from initial seeds, can be used to estimate the change in the M$_{BH}-\sigma$ relation at high redshift. We follow the reasoning presented in  \cite{Volonteri2009}: (\textit{i}) SMBHs originating from massive seeds start off above the  M$_{BH}-\sigma$ relation observed in the local universe, and migrate onto it by initially growing $\sigma$  after which further major mergers trigger accretion episodes which results in growth spurts for the MBHs; and (\textit{ii}) SMBHs originating from light seeds migrate to the local M$_{BH}-\sigma$ from below, with the MBH seeds growing without significantly altering $\sigma$. The modification of the  M$_{BH}-v_{circ}^{max}$ relation is introduced by adding a model-dependent mass scaling parameter X$(z)=\frac{\mathrm{M}_{BH}^z}{\mathrm{M}_{BH}^{z=0}}$. 
Overall, the M$_{BH}-\sigma$ relation observed at the local universe is modified to an \textit{X}$(z)$M$_{BH}^{z=0}-\frac{v_{circ}^{max}}{\eta}$ at high redshift.
Guided by tree merger simulation results \citep{Volonteri2009}, we assume a functional form of \textit{X}$=e^{b(z-2)}$, and so the redshift-dependent relation settles to the observed M$_{BH}-\sigma$ relation in the local universe at $z=2$, see left panel in Fig. \ref{fig:MV_Relation}. We allow different realizations of the parameter \textit{X}, with $b$ varying from $-0.3$ to $0.3$. For $b=0$ there is no change in the local M$_{BH}-\sigma$ relation (the mass scaling parameter, luminosity scaling parameter, and duty cycle are correlated with the SMBH mass at a later stage of its evolution, and we take this into account during our MC simulation; see section 2.4). 
A model with positive (negative) value of the parameter $b$ implies that the MBH that resides at the centre of a DM halo with a specific value of $v_{circ}^{max}$ is heavier (lighter) than the MBH found at lower redshift in a DM halo of similar $v_{circ}^{max}$, \textit{i.e.,} a steepening (flatenning) of the slope of the M$_{BH}-\sigma$ relation at high redshifts; see right panel in Fig. \ref{fig:MV_Relation}. We find for each specific model what is the minimum $v_{circ}^{max}$ for which a halo will host a MBH that can be detected with each telescope. The value of min$(v_{circ}^{max})$ will be used next to estimate the density of observed objects using halo densities derived from cosmological simulations of the evolution of the large-scale structure in the universe.  
\subsection{Deriving the density of observed sources}
Once we determine the min($v_{circ}^{max}$) of an halo that hosts a detectable MBH for each model, we ask what is the comoving density of halos with $v^{max}_{circ}>min(v^{max}_{circ})$. This allows us to estimate the spatial density of detectable MBHs from each model with each telescope.\\
In this work we use the Bolshoi N-body simulation\footnote{\textit{hipacc.ucsc.edu/Bolshoi}} to estimate the density of MBHs at high redshifts based on the modified M$_{BH}-\sigma$ relation and min($v_{circ}^{max}$) for a detectable MBH. The Bolshoi simulation \citep{Klypin2011} covers a volume of $250\,h^{-1}\,$Mpc on a side using $\sim8\,$billion particles, allowing to probe halos with $v_{circ}^{max}$ as low as $50\,\mathrm{km\,s^{-1}}$. It uses $v_{circ}^{max}$ as the main property of halos, since it is a more stable quantity for characterizing physical parameters of the central regions of dark matter halos than the halo virial mass - and therefore a better quantity for relating dark matter halos to the galaxies they host. \cite{Klypin2011} show that $90\%$ of the halos have circular velocities within $8\%$ of the median value. Using $v_{circ}^{max}$ as the main characteristic of a DM halo, we directly derive the cumulative MBH density function using the modified M$_{BH}-\sigma$ relation.\\
Over a broad range in mass and redshift, the halo cumulative number density can be parametrized according to \citep{Klypin2011}:
$$
n(>v)=Av^{-3}exp(-[\frac{v}{v_0}]^\alpha)
$$
Where $v$ is the halo velocity.
The parameters in the density function, $A$, $\alpha$, and $v_0$, are evolving according to $\sigma_8(z)$, the variance in the density perturbation amplitude smoothed over an $8{\it{h}}^{-1}$Mpc box, with\footnote{We assume $\sigma_8$ evolves according to the scale factor $a(t)$ and is normalized to 0.82, in agreement with recent Planck results \citep{Planck2016}.}:
\begin{equation}
A=1.52\times10^4\sigma^{-3/4}_8(z) [h^{-1}\,\mathrm{Mpc/km\,s^{-1}}],
\end{equation}
\begin{equation}
\alpha=1+2.15\sigma_8^{4/3}(z), 
\end{equation}
\begin{equation}
v_0=3300\frac{\sigma^2_8(z)}{1+2.5\sigma_8^2(z)} [\mathrm{km/s}].
\end{equation}	
By integrating the comoving halo density equation with a lower bound set to min($v_{circ}^{max}$), we get the number of detectable MBHs for each model $N_{Obs}^\dagger$. 
\begin{equation}
N_{Obs}^\dagger(z)=\int_{z'=z-\Delta z}^{z'=z+\Delta z} \left[\frac{dV}{dz}(d\Omega) \times  n(z';v^{max}_{circ}>min(v^{max}_{circ})\right] dz'
\end{equation}
In the equation above we assume each halo has a MBH at its center (\textit{i.e.}, the occupancy $\eta_{oc}=1$), and that each MBH in a given FoV is continuously accreting and radiating at the specified value of $\lambda$ (\textit{i.e.}, the duty cycle $\eta_{dc}=1$). The actual number of observed objects is $N_{obs}(z)=\eta_{oc}\times \eta_{dc}\times N_{Obs}^\dagger(z)$. Table 1 summaries the parameter space we explore in this paper.
\subsection{Monte Carlo Realization}
\begin{figure}[]
\begin{center}
{\includegraphics[width=0.45\columnwidth]{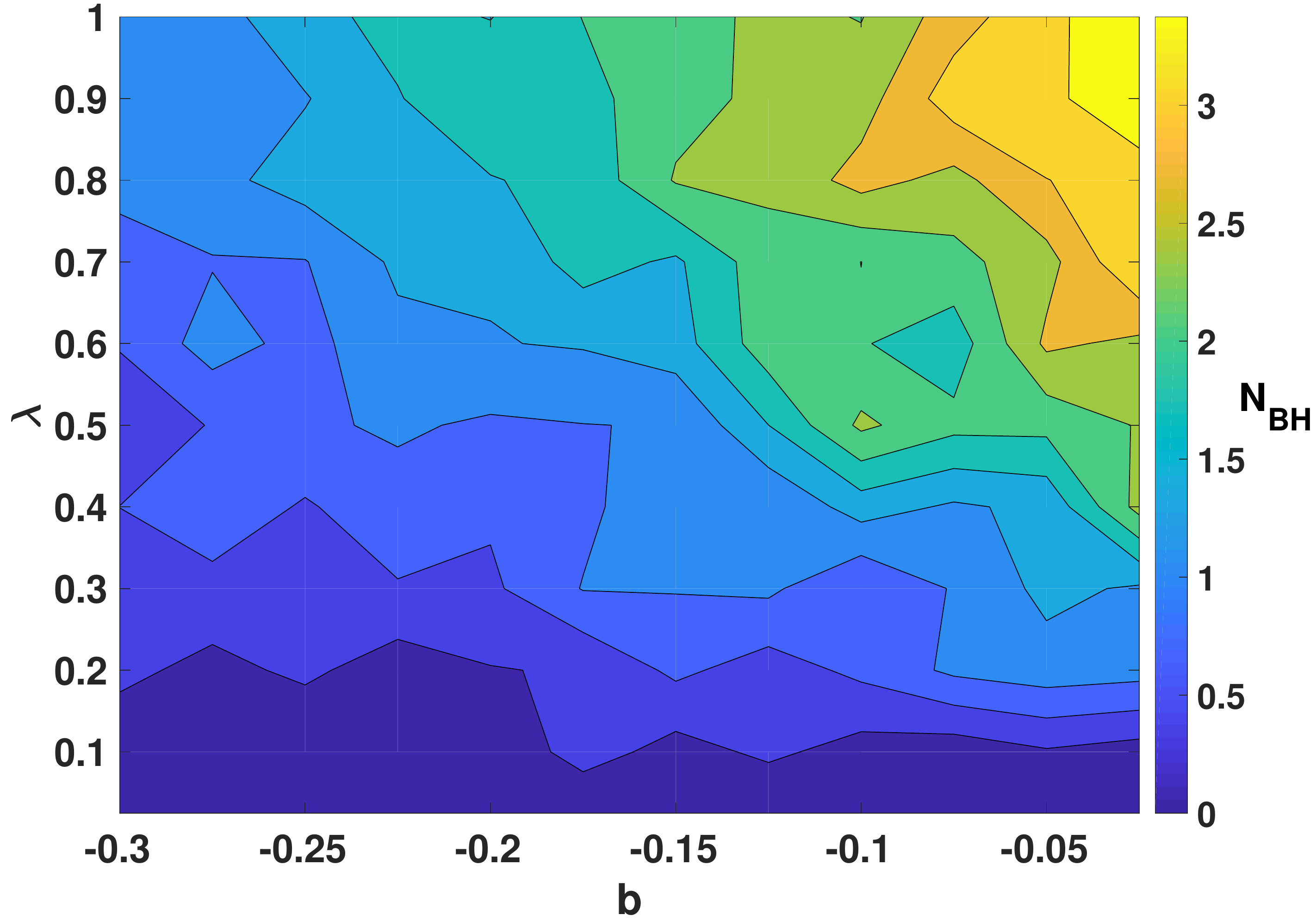}} 
{\includegraphics[width=0.45\columnwidth]{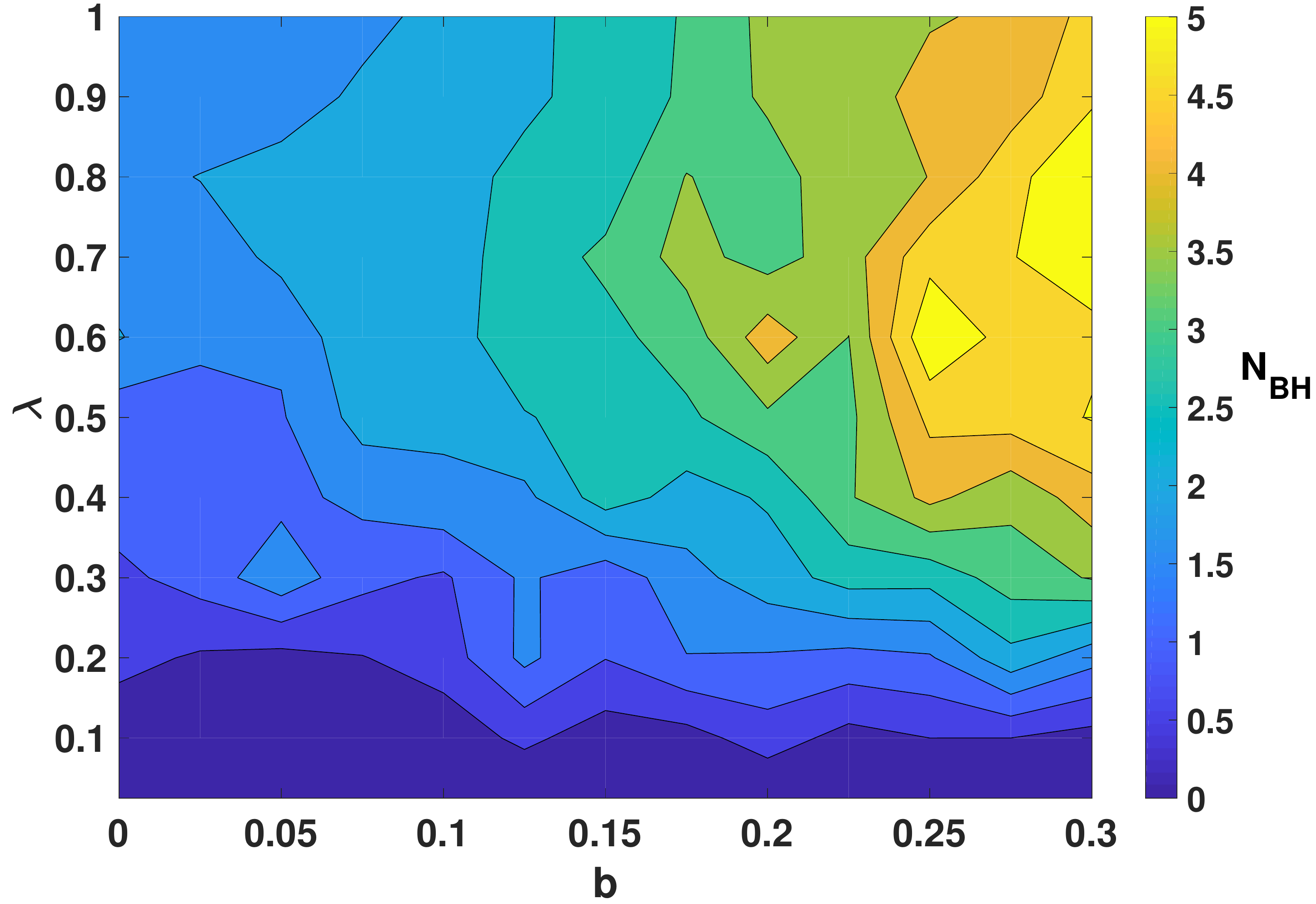}} \\
{\includegraphics[width=0.45\columnwidth]{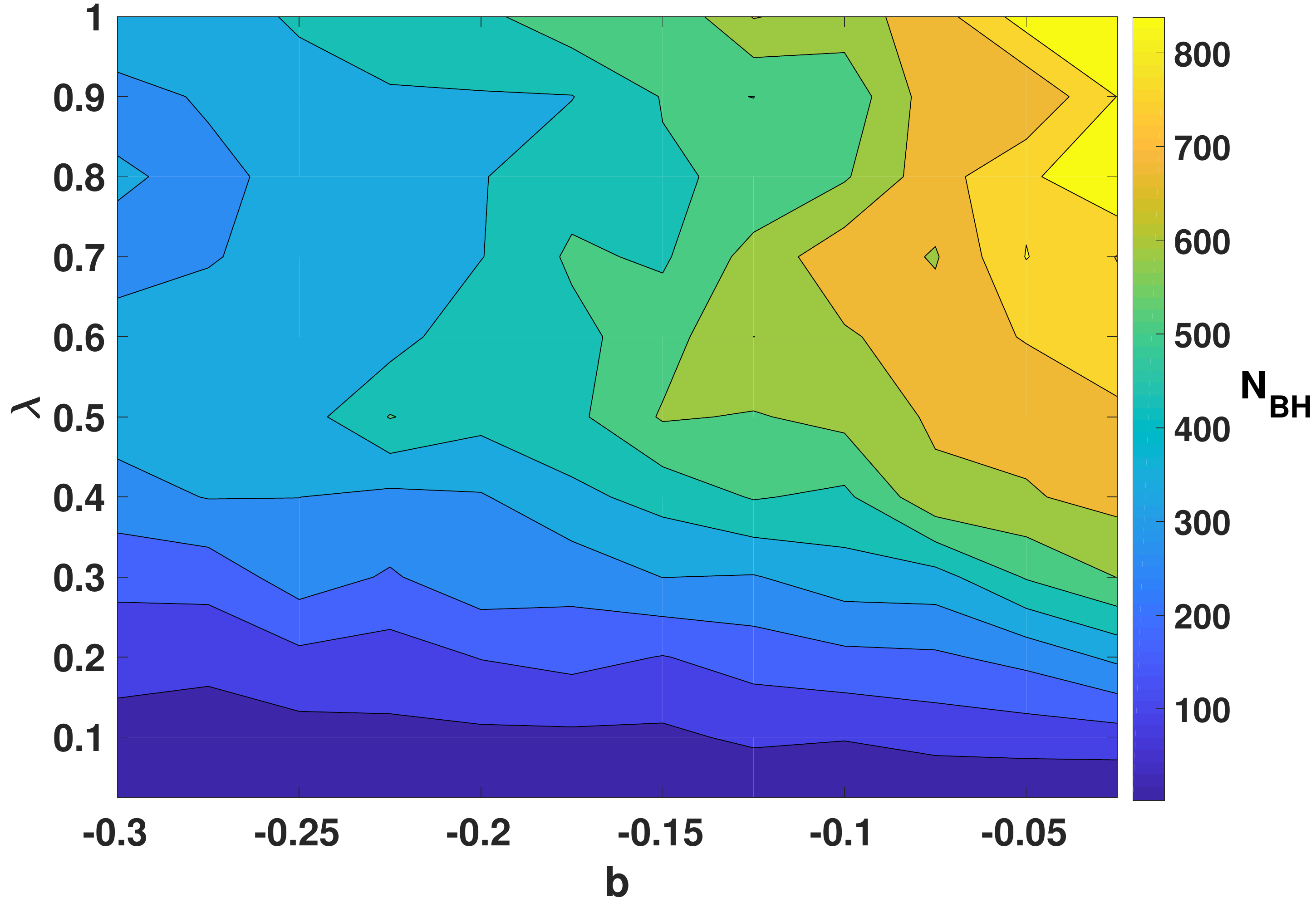}} 
{\includegraphics[width=0.45\columnwidth]{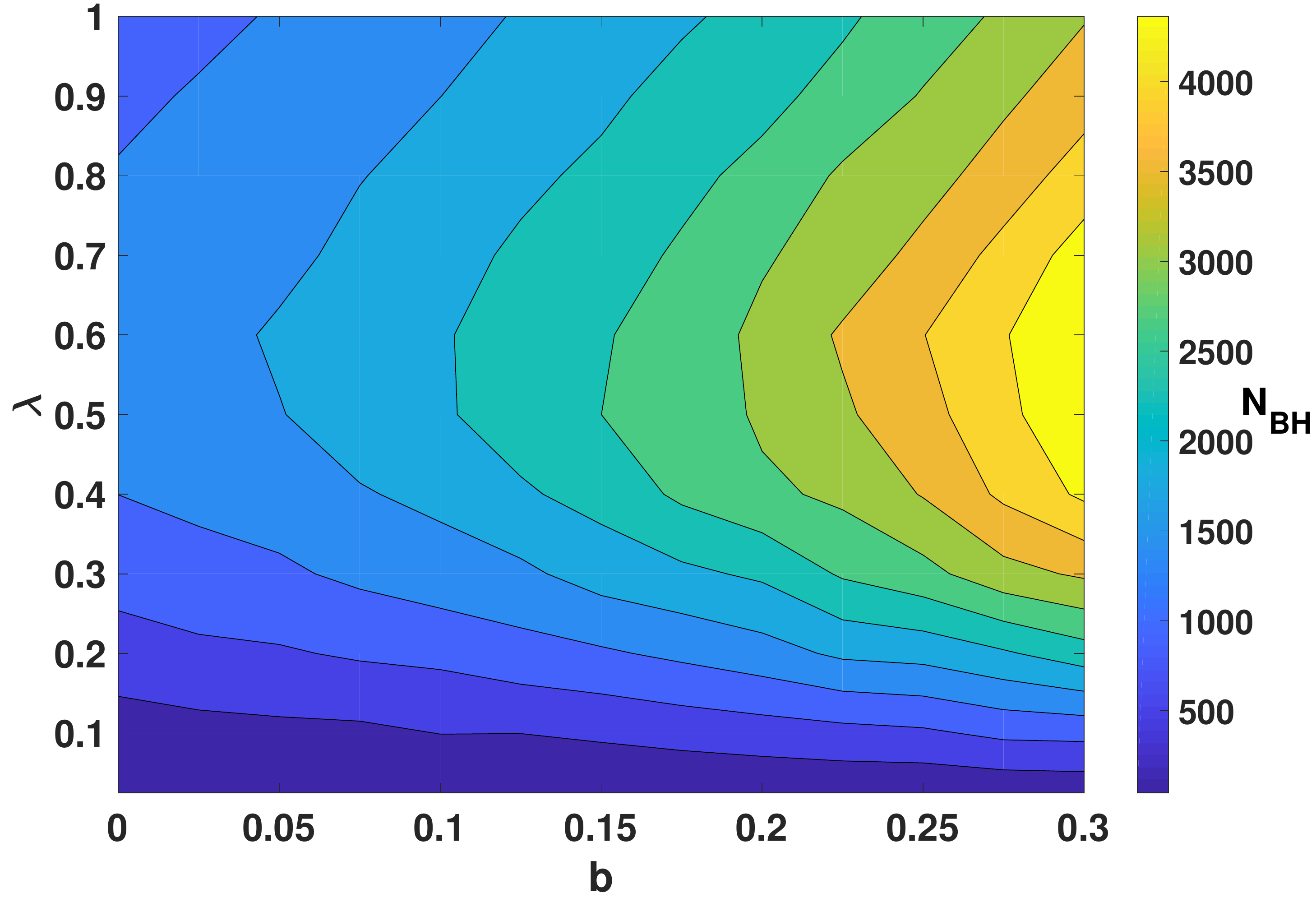}} 
\end{center}
\caption{Expected number of observed MBHs for Chandra (top) and Lynx (bottom) as a function of the luminosity scaling parameter, $\lambda$, and the mass scaling parameter, b. Light (Heavy) seed models appear on the left (right) panels. While Chandra is expected to detected a few MBHs at most, Lynx is expected to detected a statistical significant sample of MBHs for both light and heavy seed models.}
\label{fig:N_bh}
\end{figure}

\begin{figure}[]
\begin{center}
{\includegraphics[width=0.22\columnwidth]{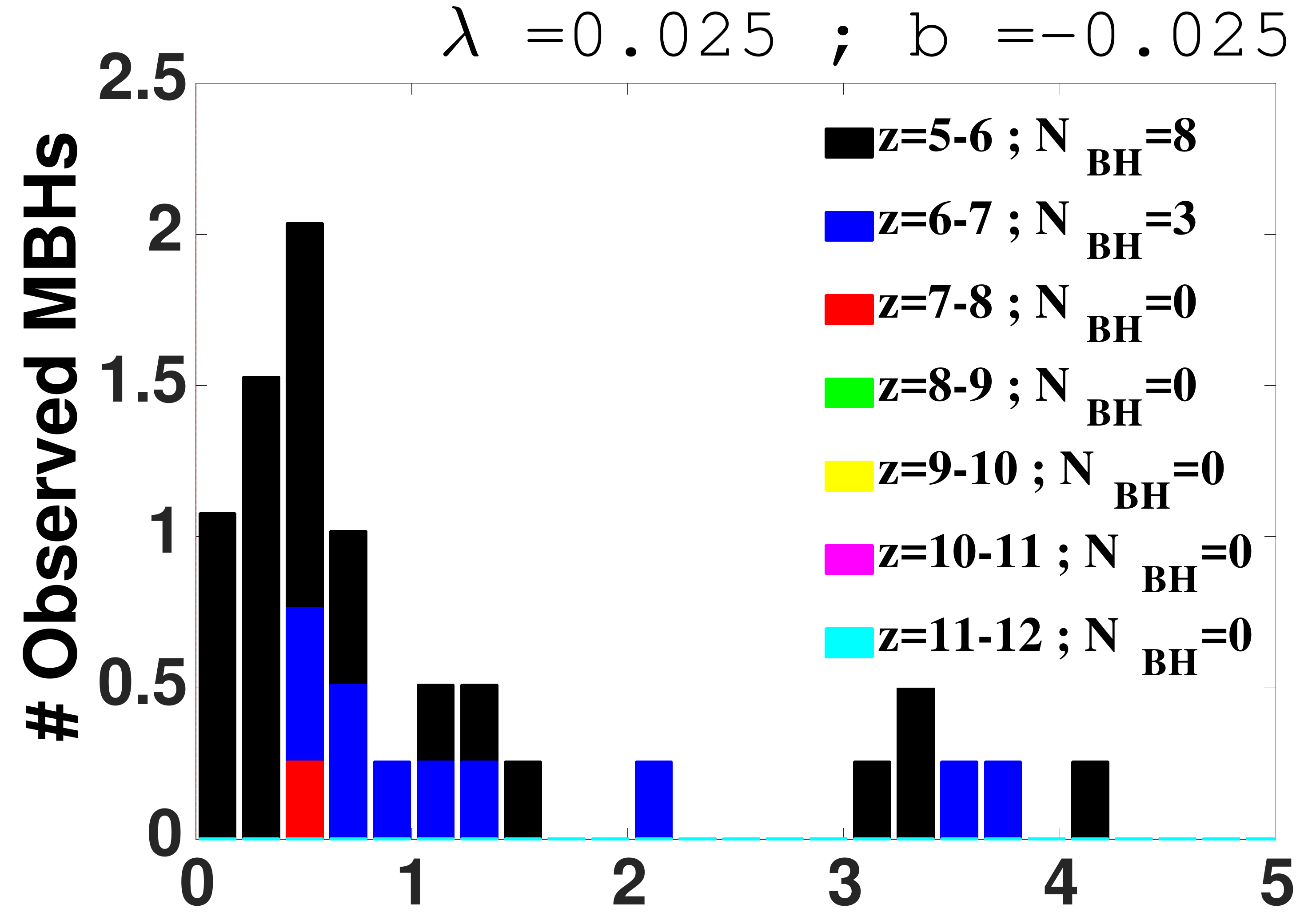}} 
{\includegraphics[width=0.22\columnwidth]{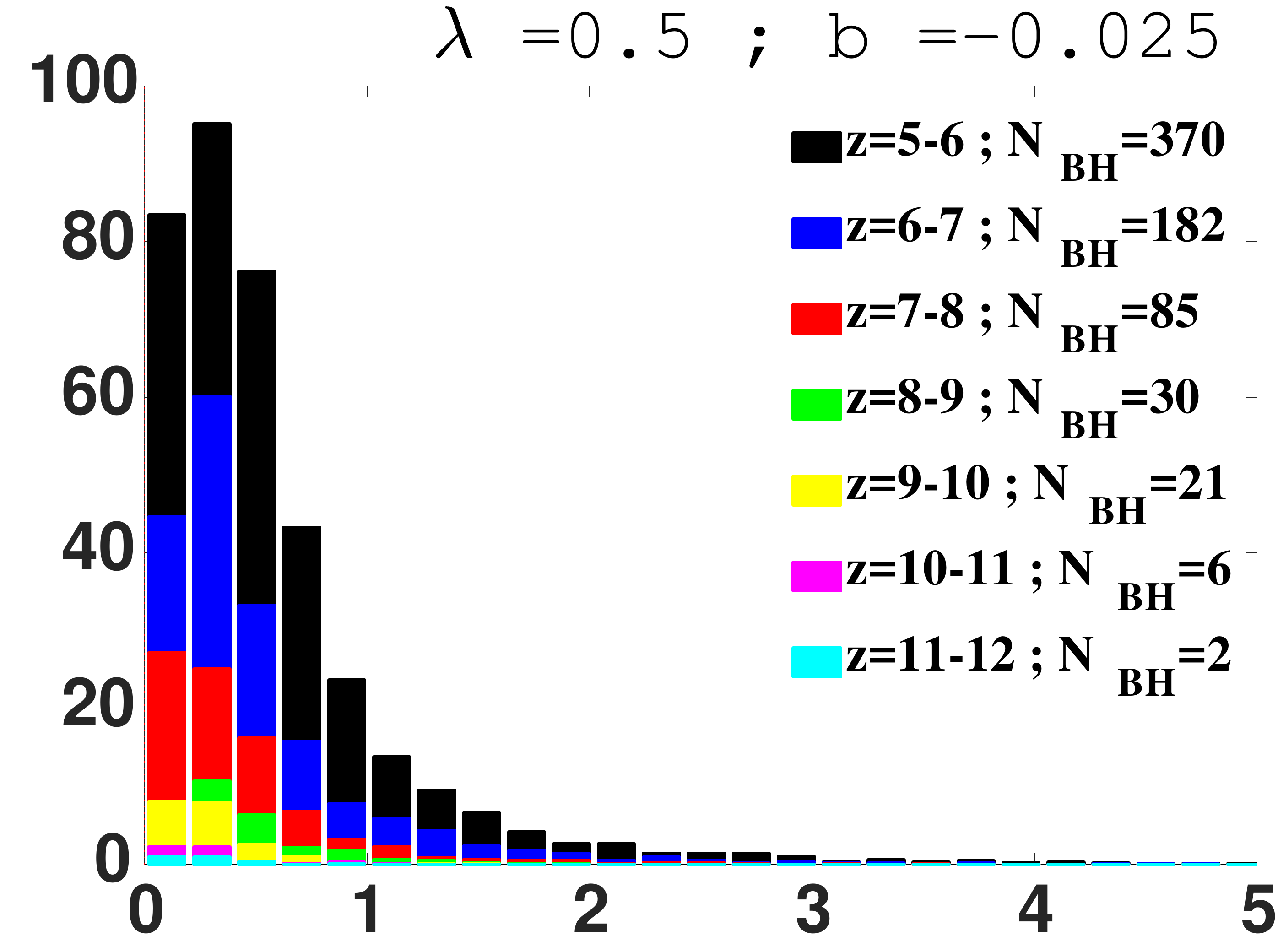}} 
{\includegraphics[width=0.22\columnwidth]{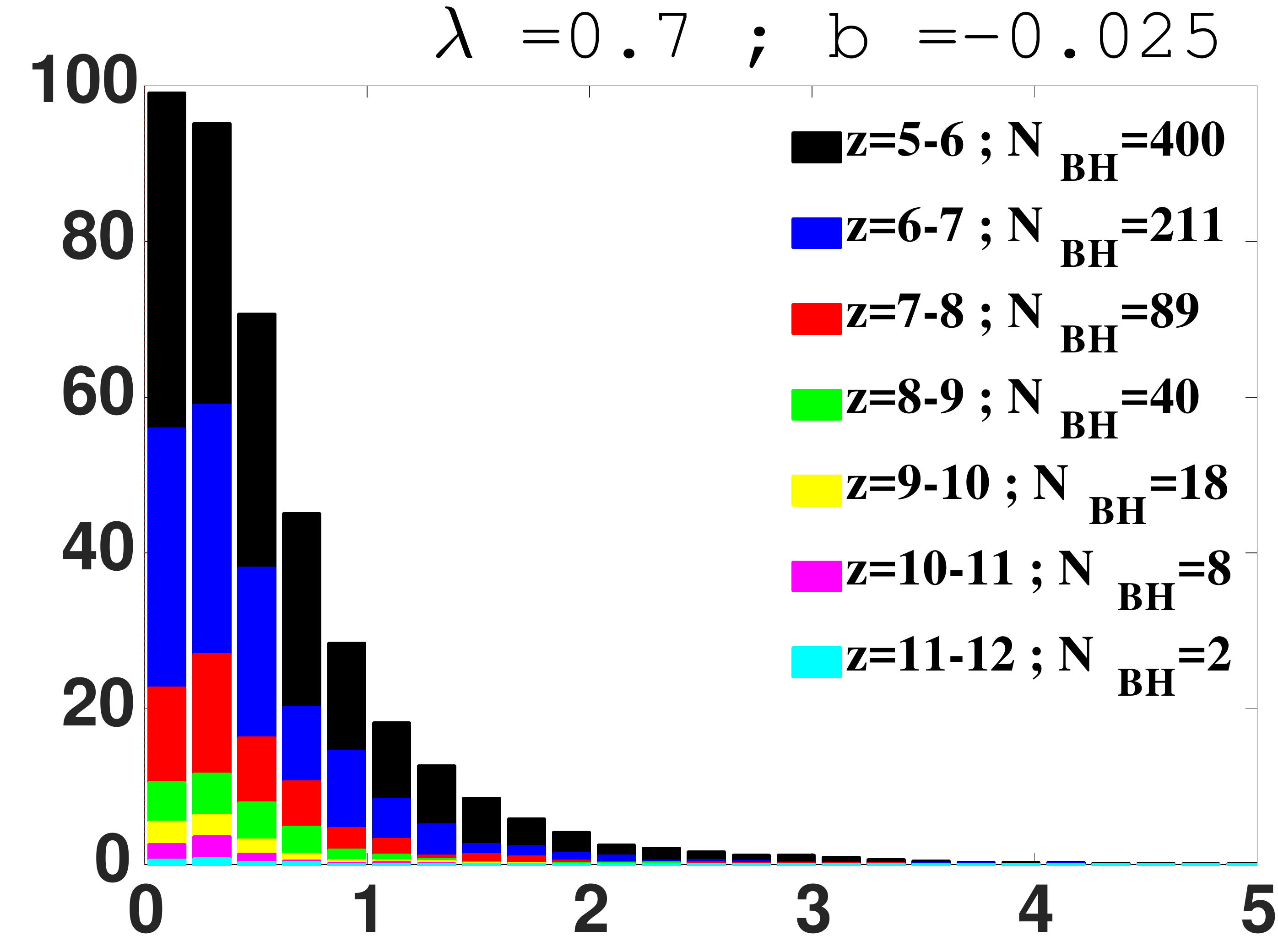}} 
{\includegraphics[width=0.22\columnwidth]{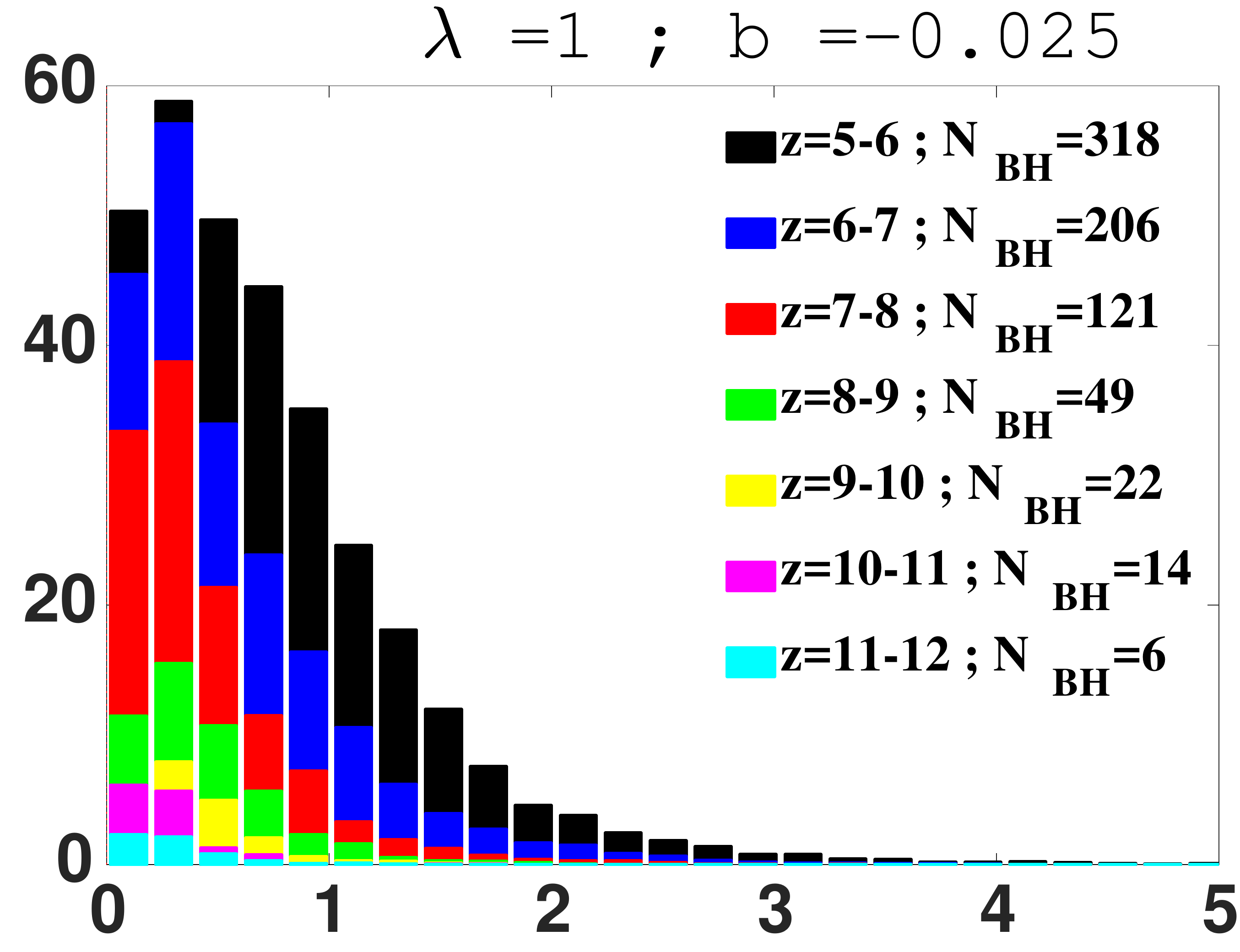}} \\
{\includegraphics[width=0.22\columnwidth]{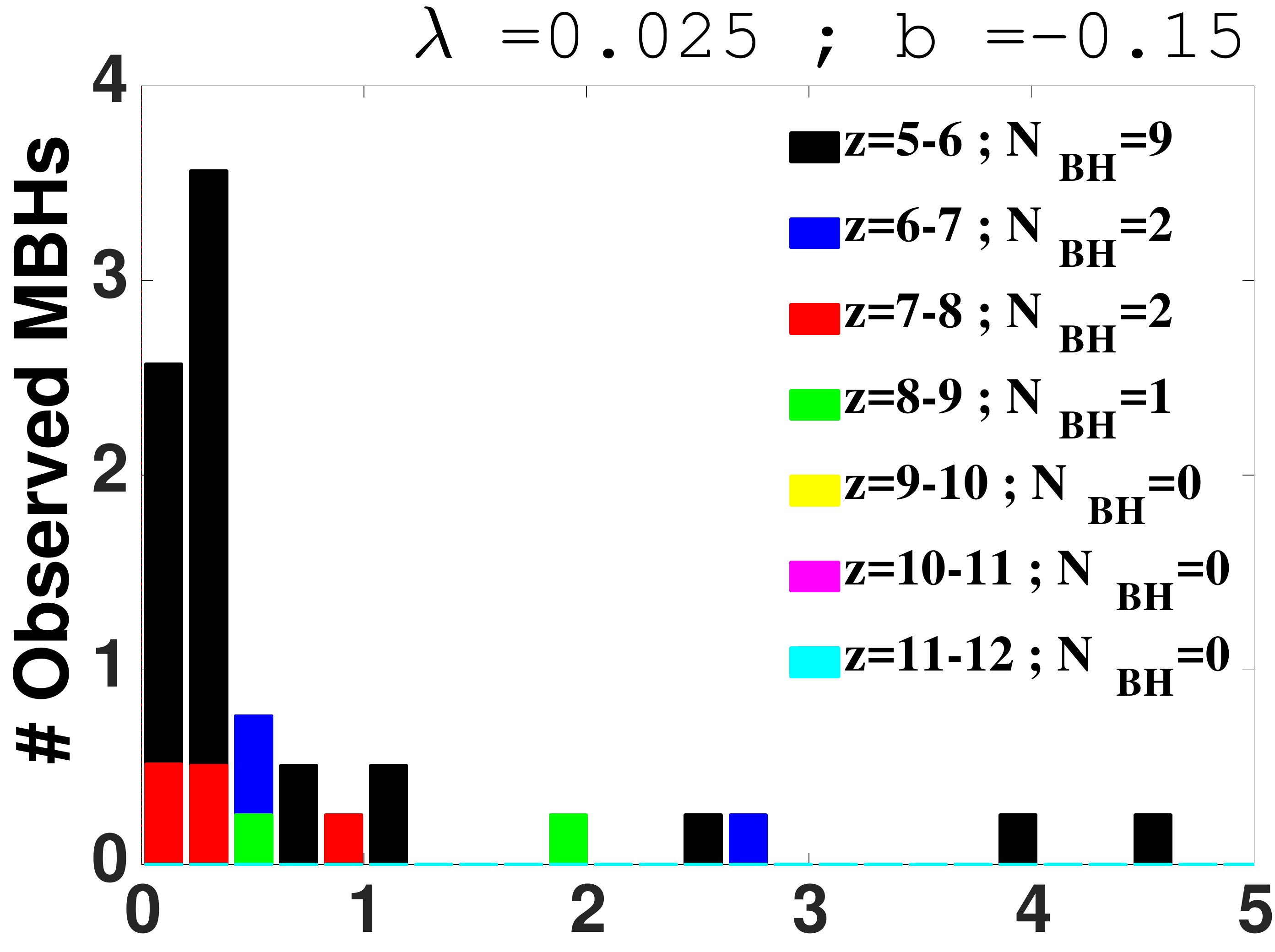}} 
{\includegraphics[width=0.22\columnwidth]{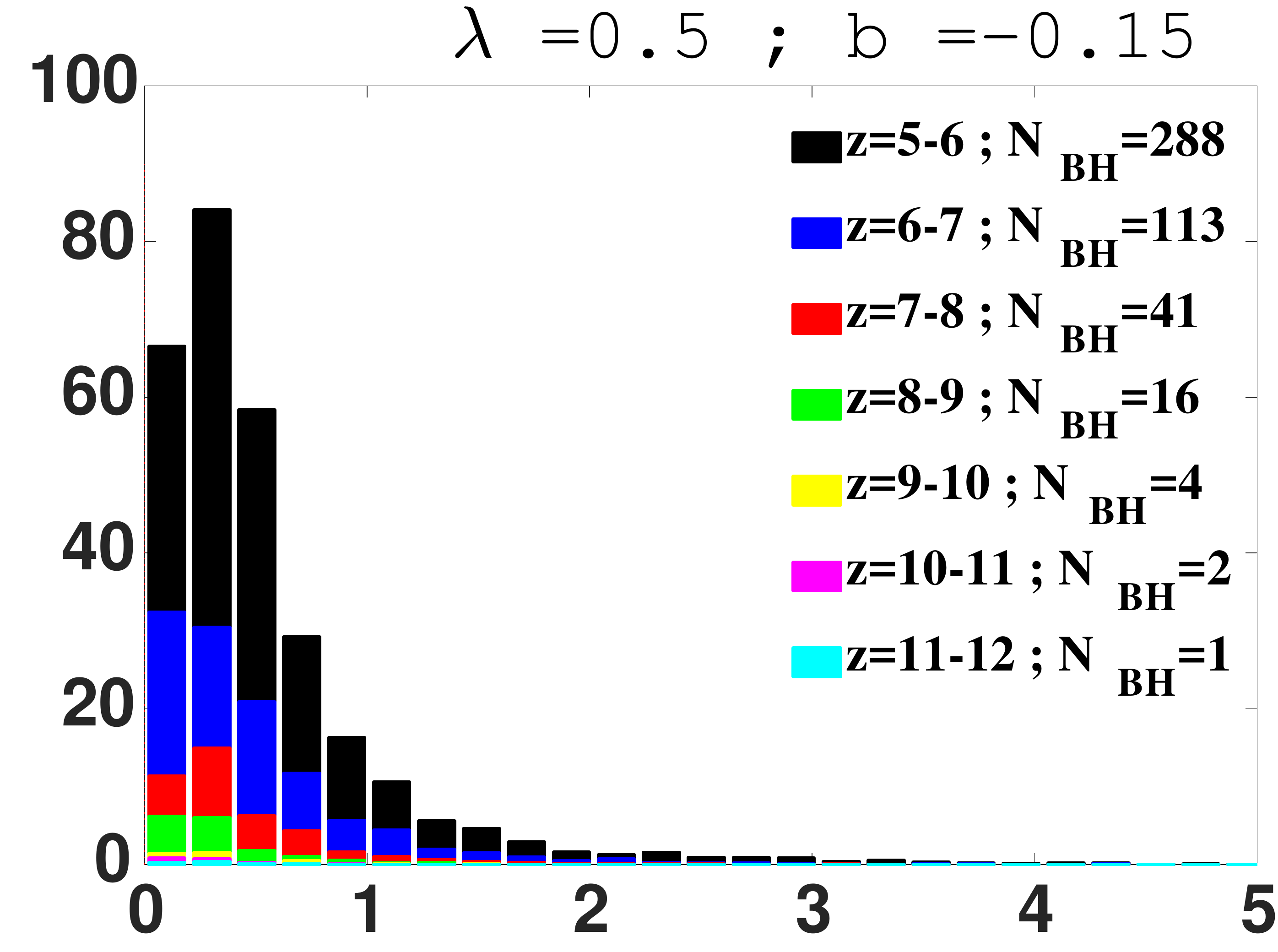}} 
{\includegraphics[width=0.22\columnwidth]{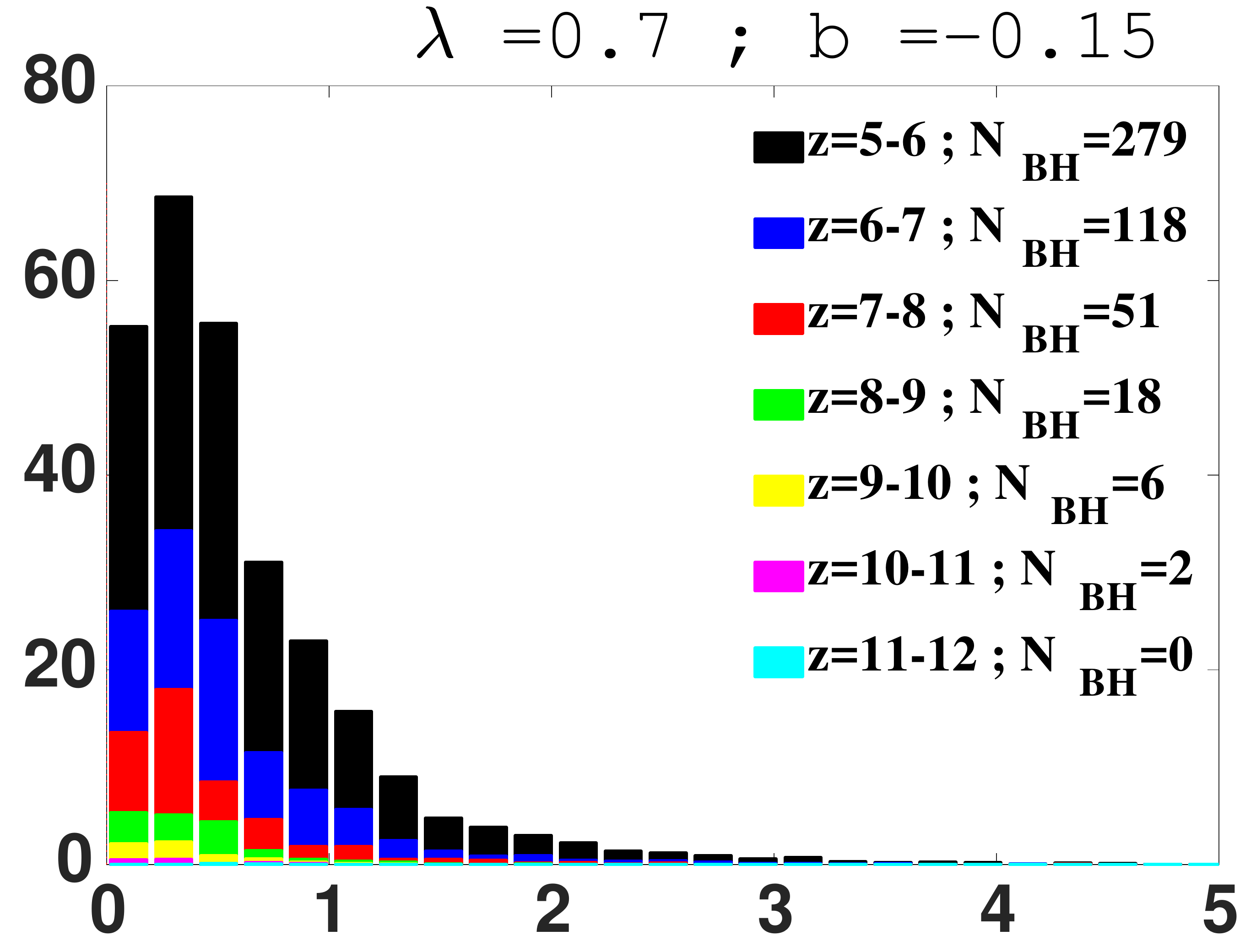}} 
{\includegraphics[width=0.22\columnwidth]{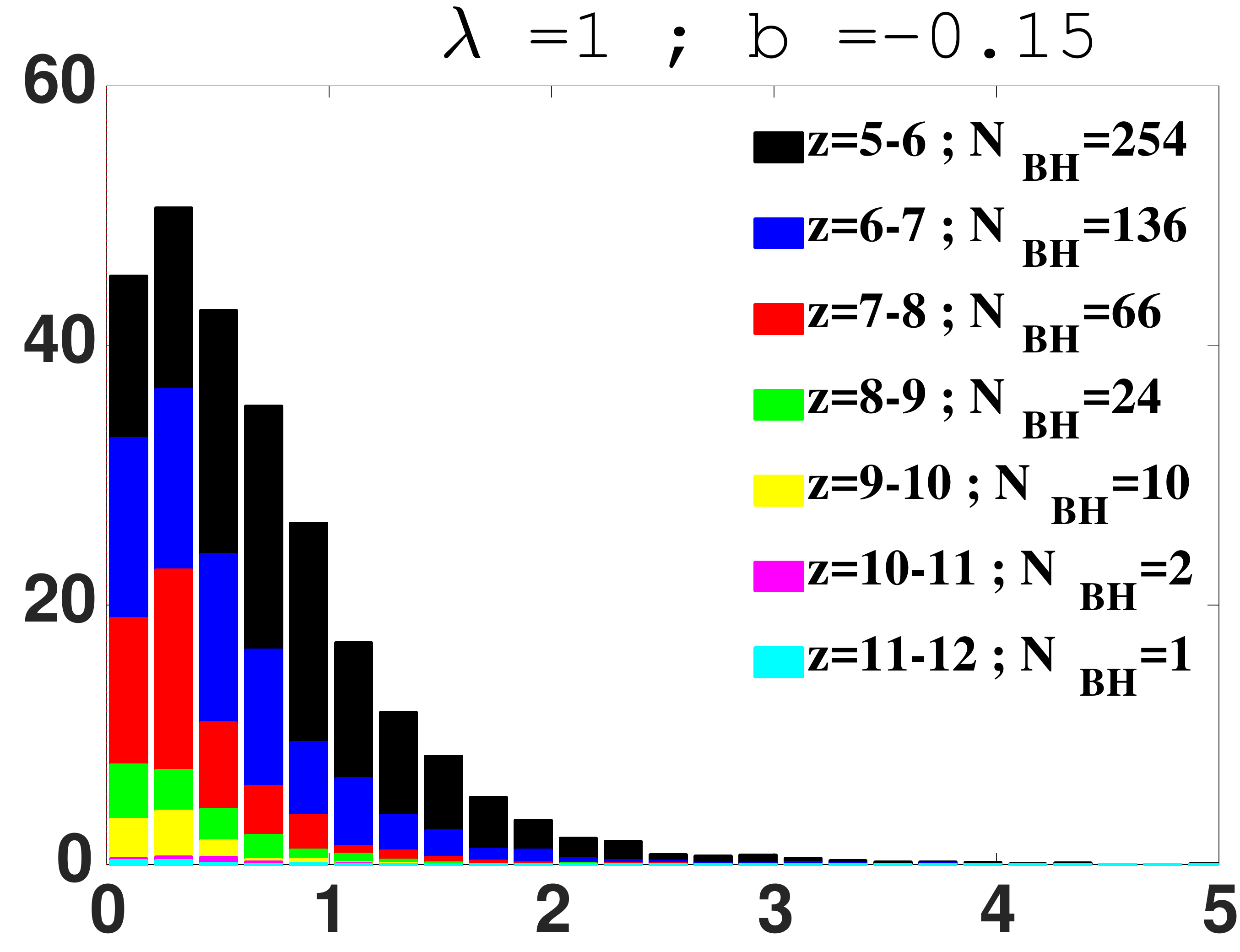}} \\
{\includegraphics[width=0.22\columnwidth]{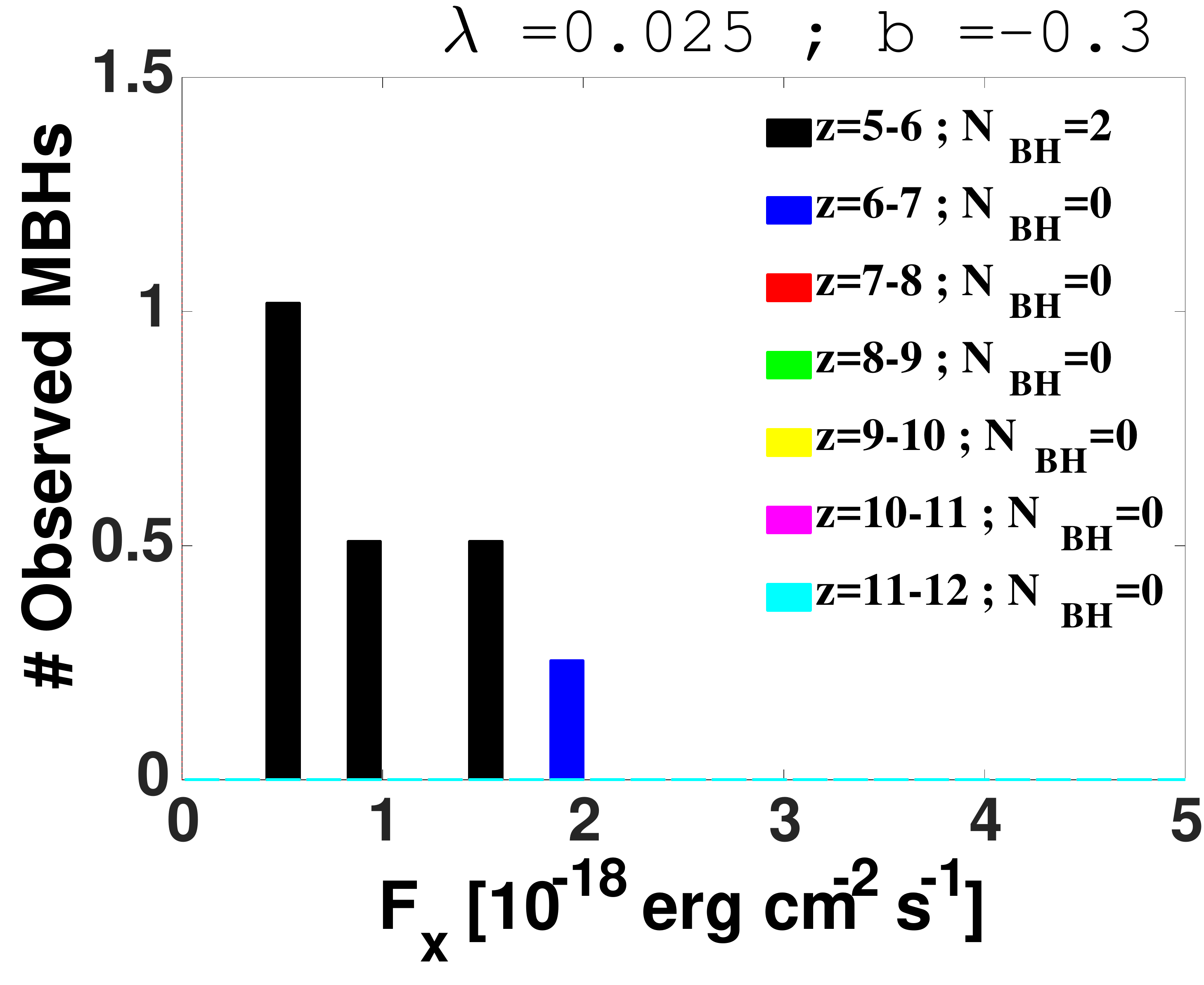}} 
{\includegraphics[width=0.22\columnwidth]{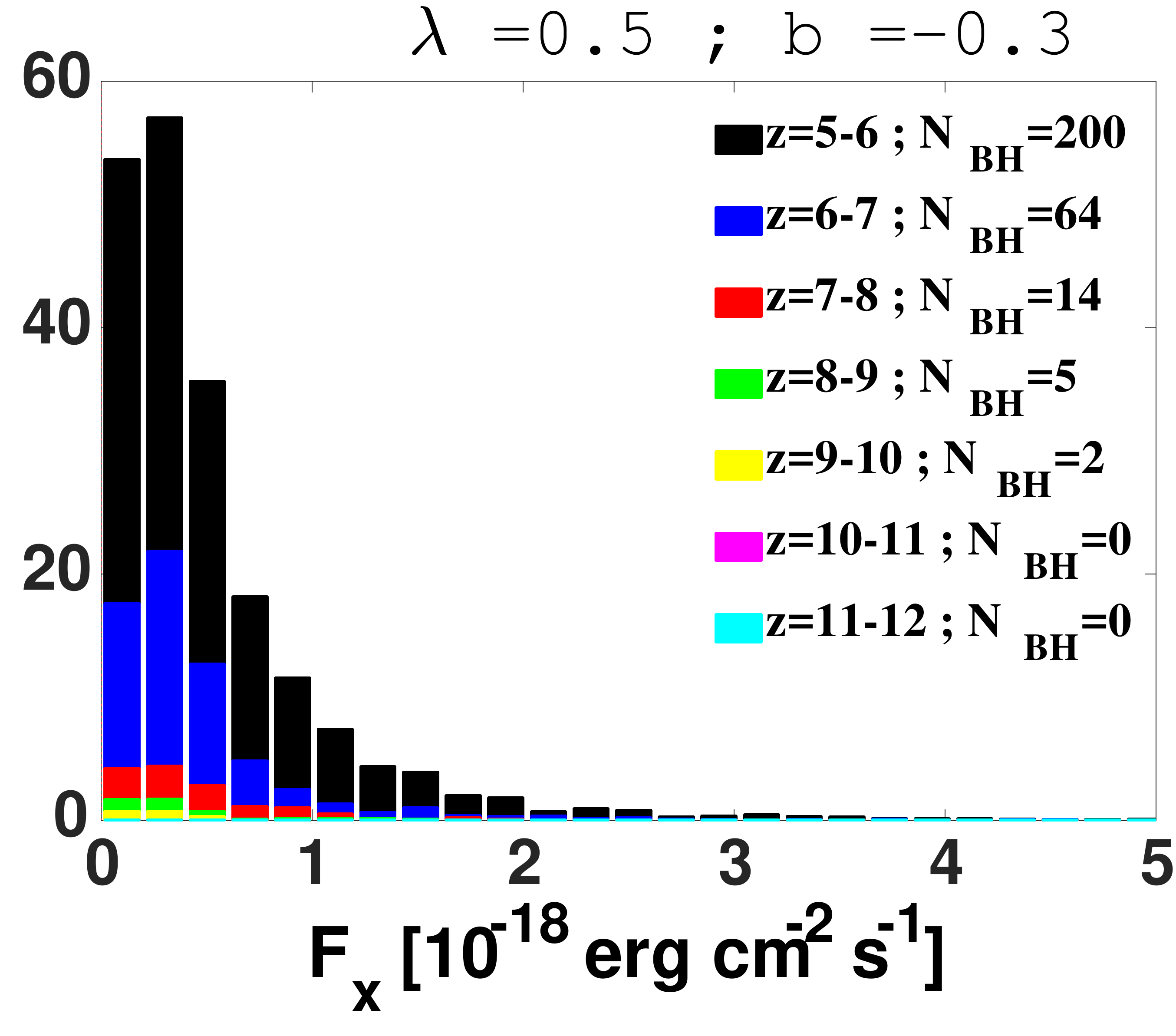}} 
{\includegraphics[width=0.22\columnwidth]{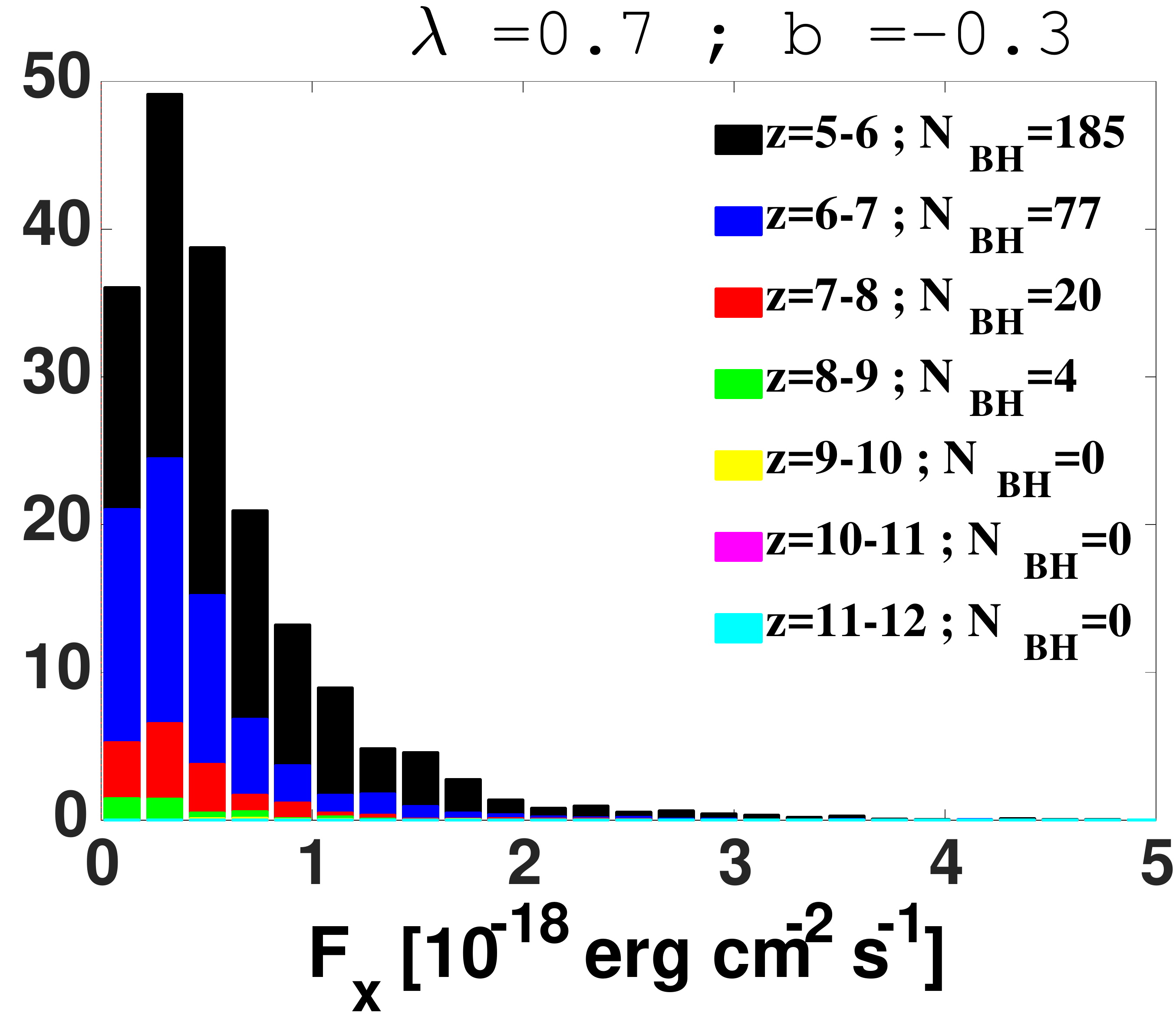}} 
{\includegraphics[width=0.22\columnwidth]{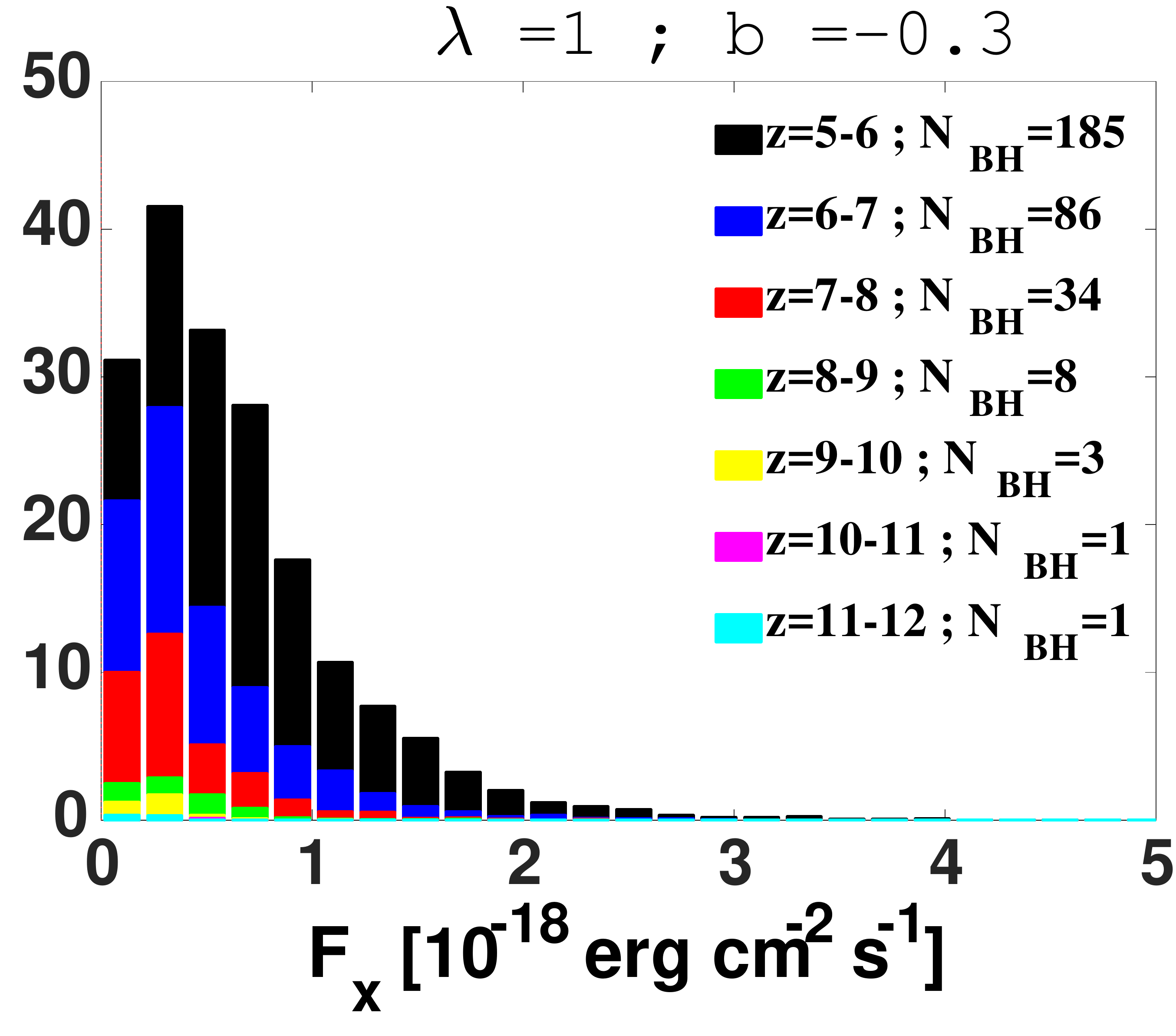}} \\
\caption{Expected X-ray flux distributions for light seed models as observed with the X-ray surveyor.}
\end{center}
\label{fig:LightFxDist}
\end{figure}

\begin{figure}[]
\begin{center}																  
{\includegraphics[width=0.22\columnwidth]{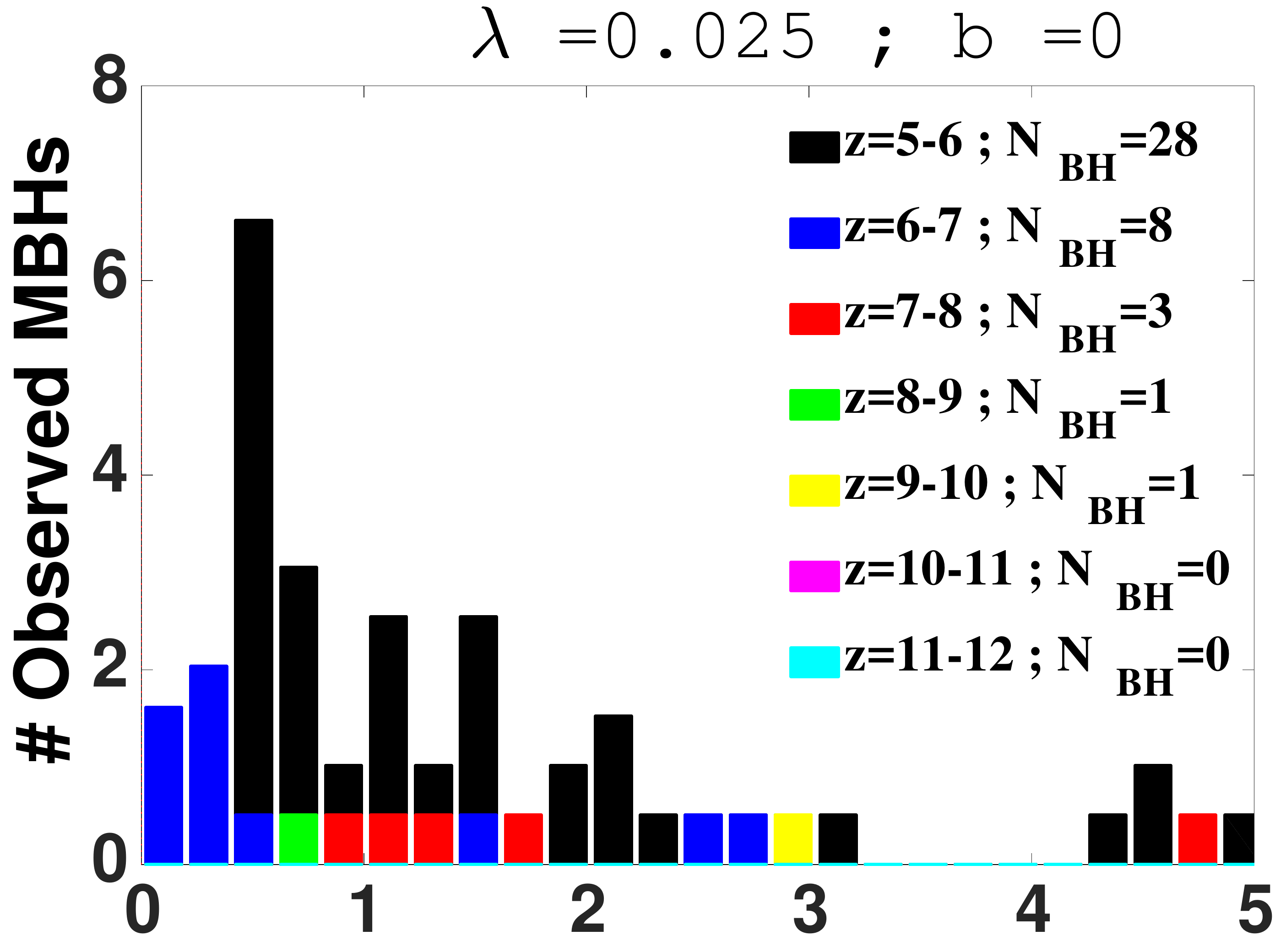}} 
{\includegraphics[width=0.22\columnwidth]{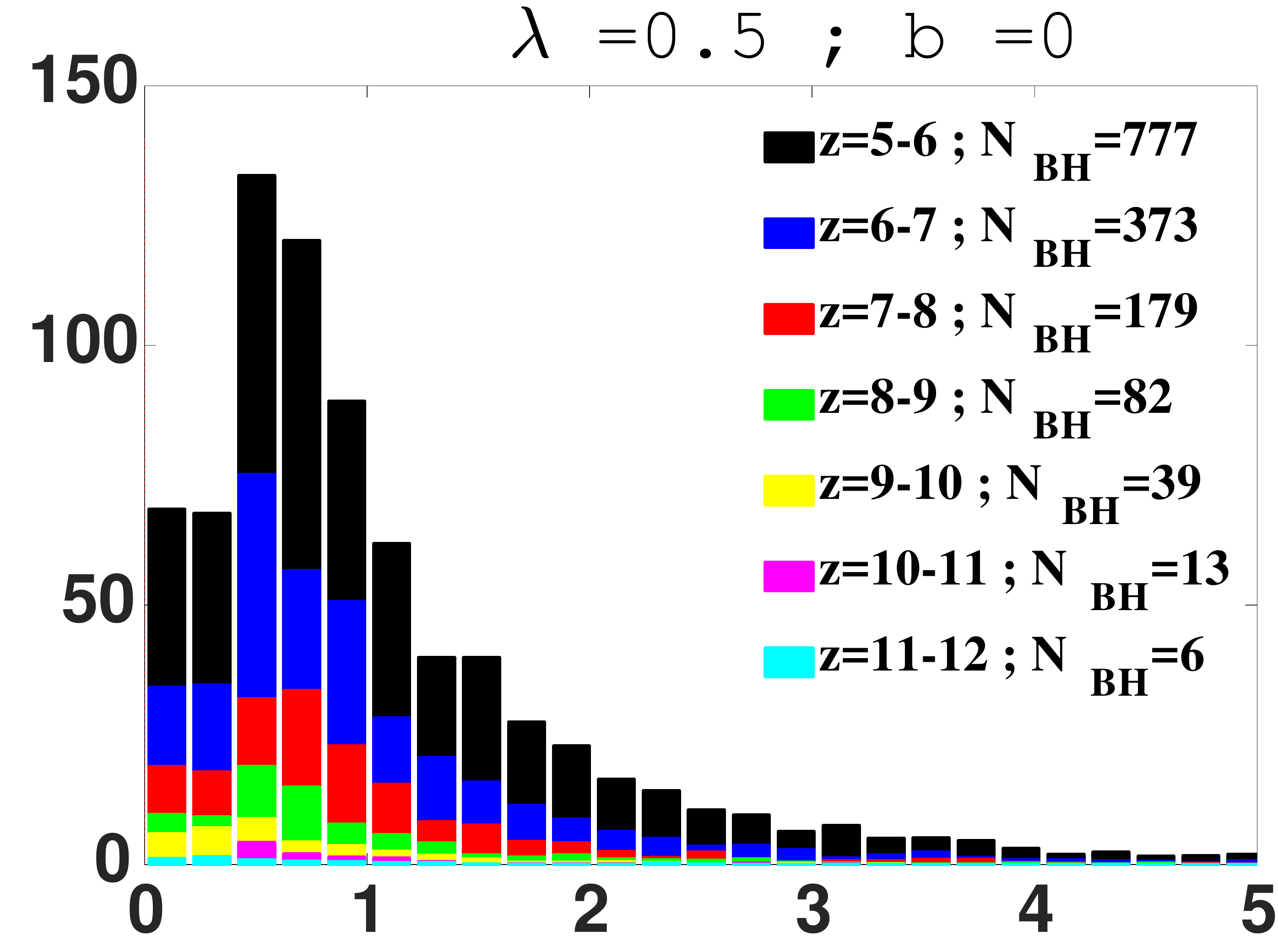}} 
{\includegraphics[width=0.22\columnwidth]{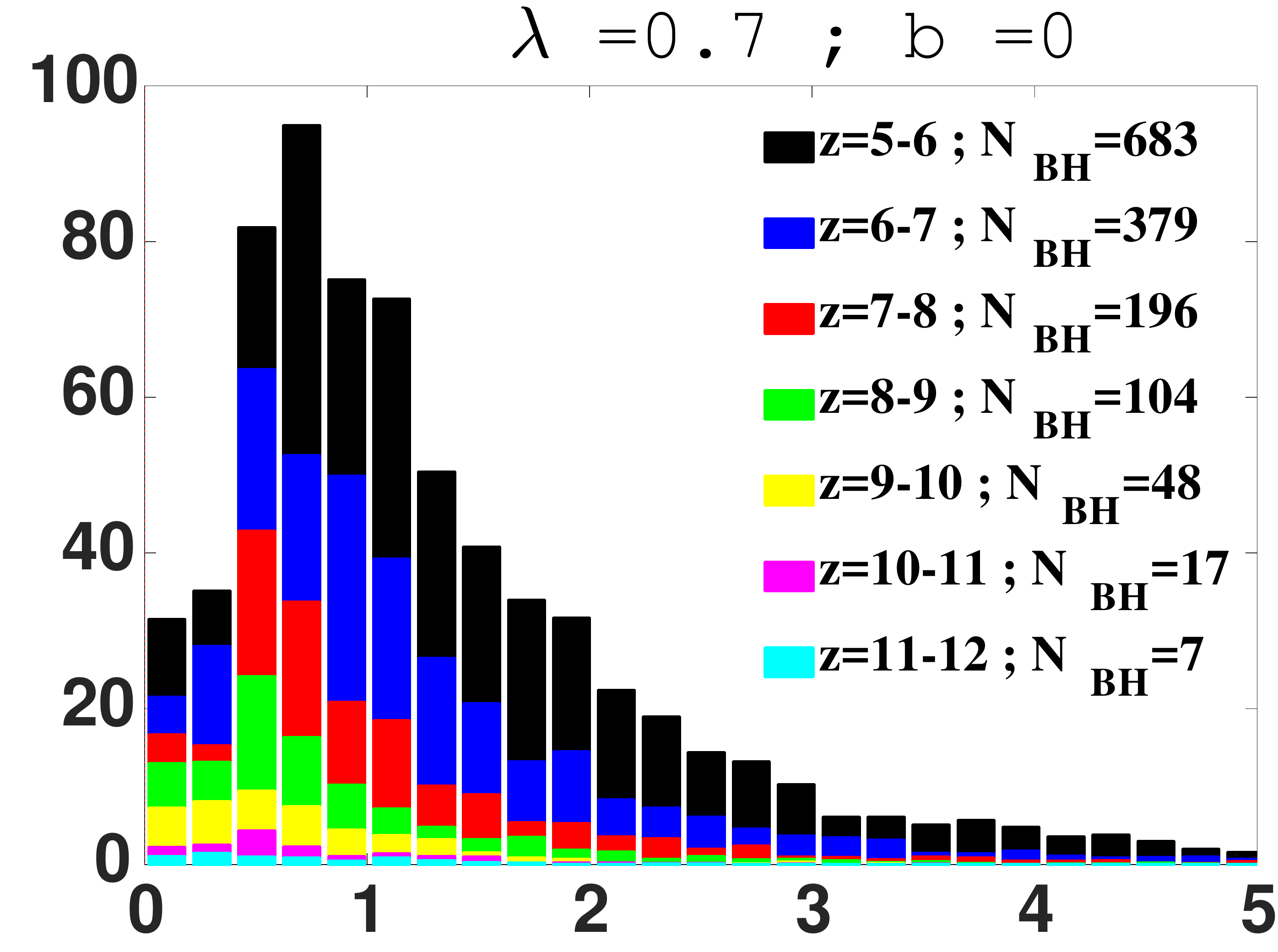}} 
{\includegraphics[width=0.22\columnwidth]{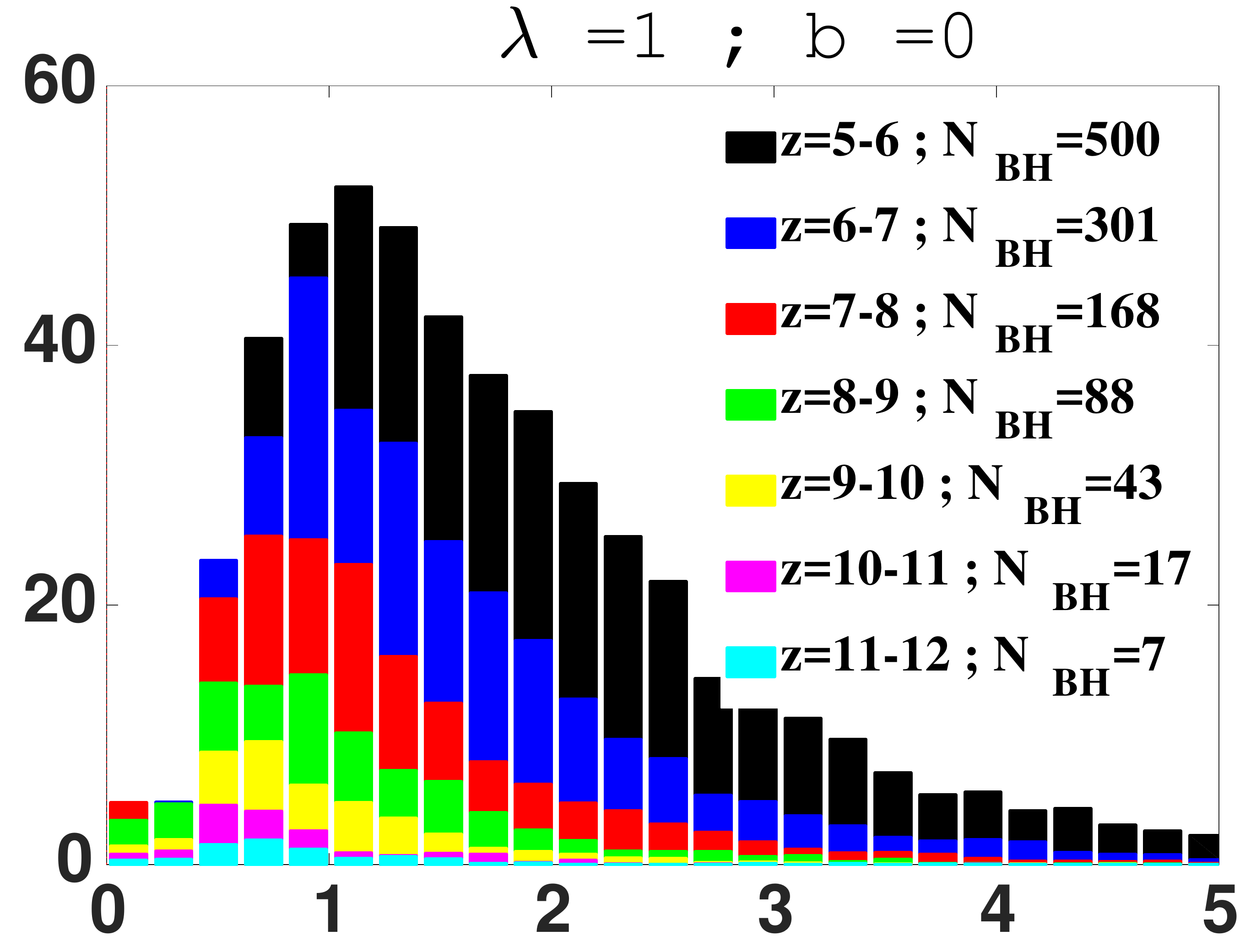}} \\
{\includegraphics[width=0.22\columnwidth]{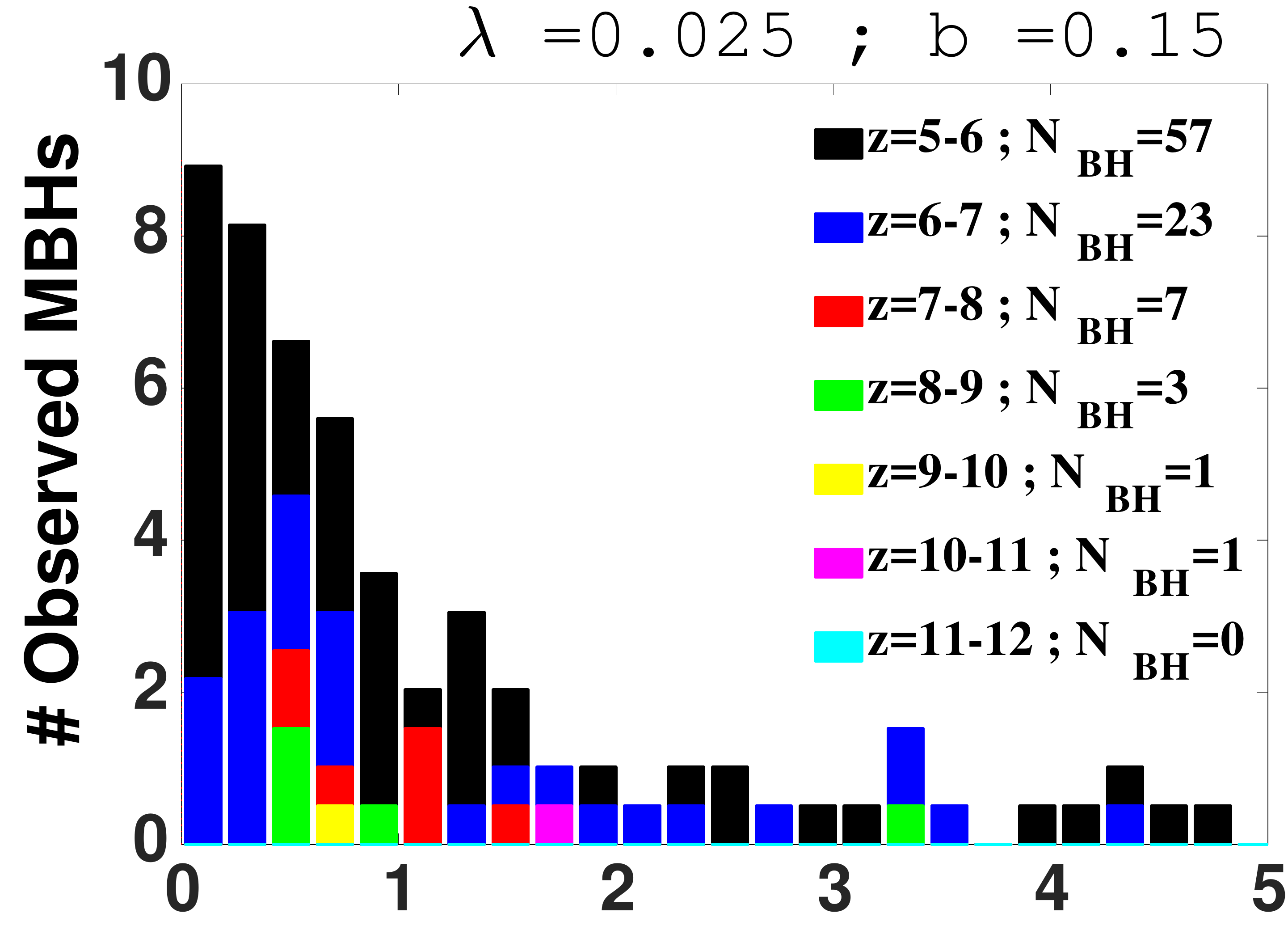}} 
{\includegraphics[width=0.22\columnwidth]{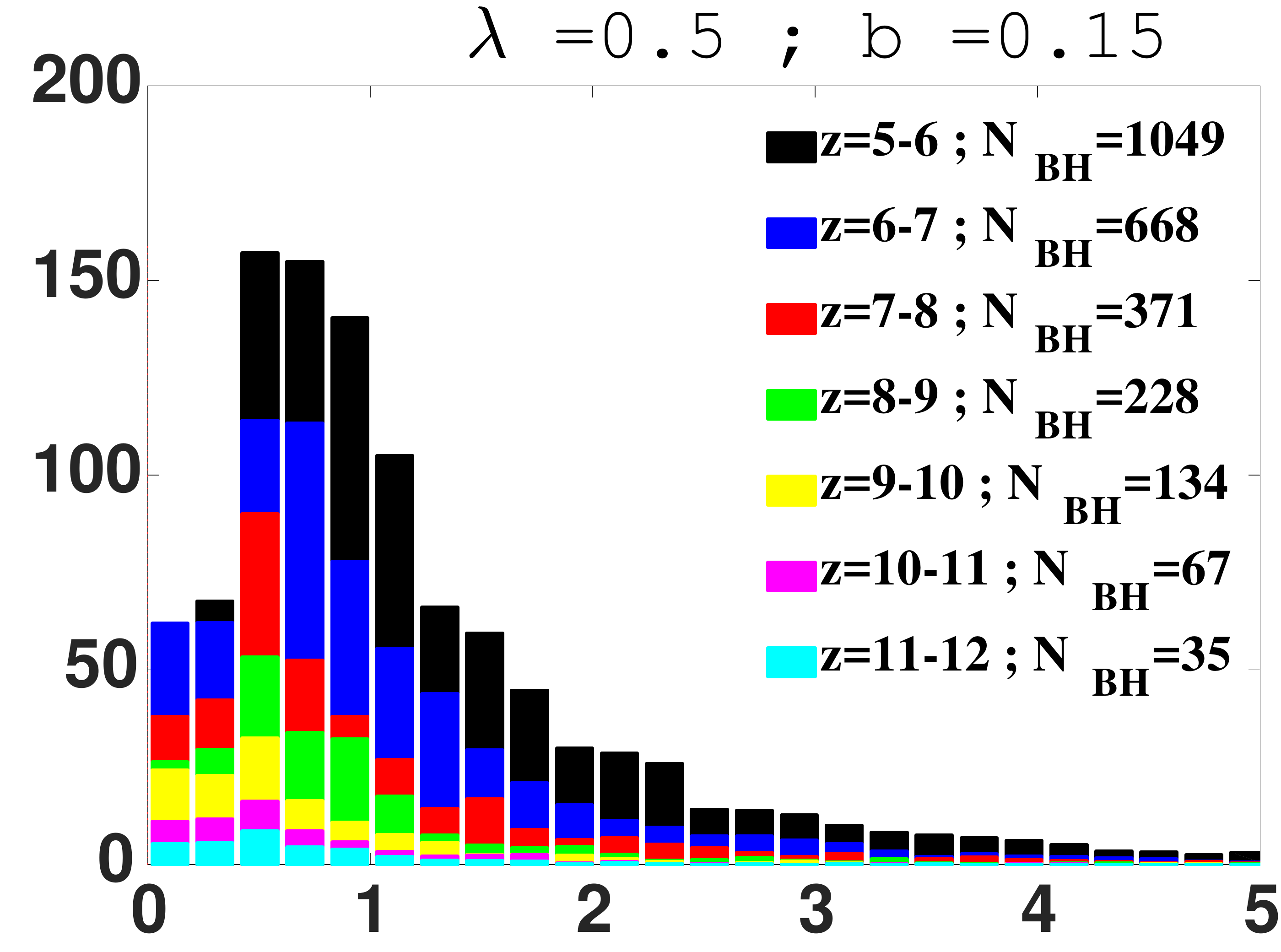}} 
{\includegraphics[width=0.22\columnwidth]{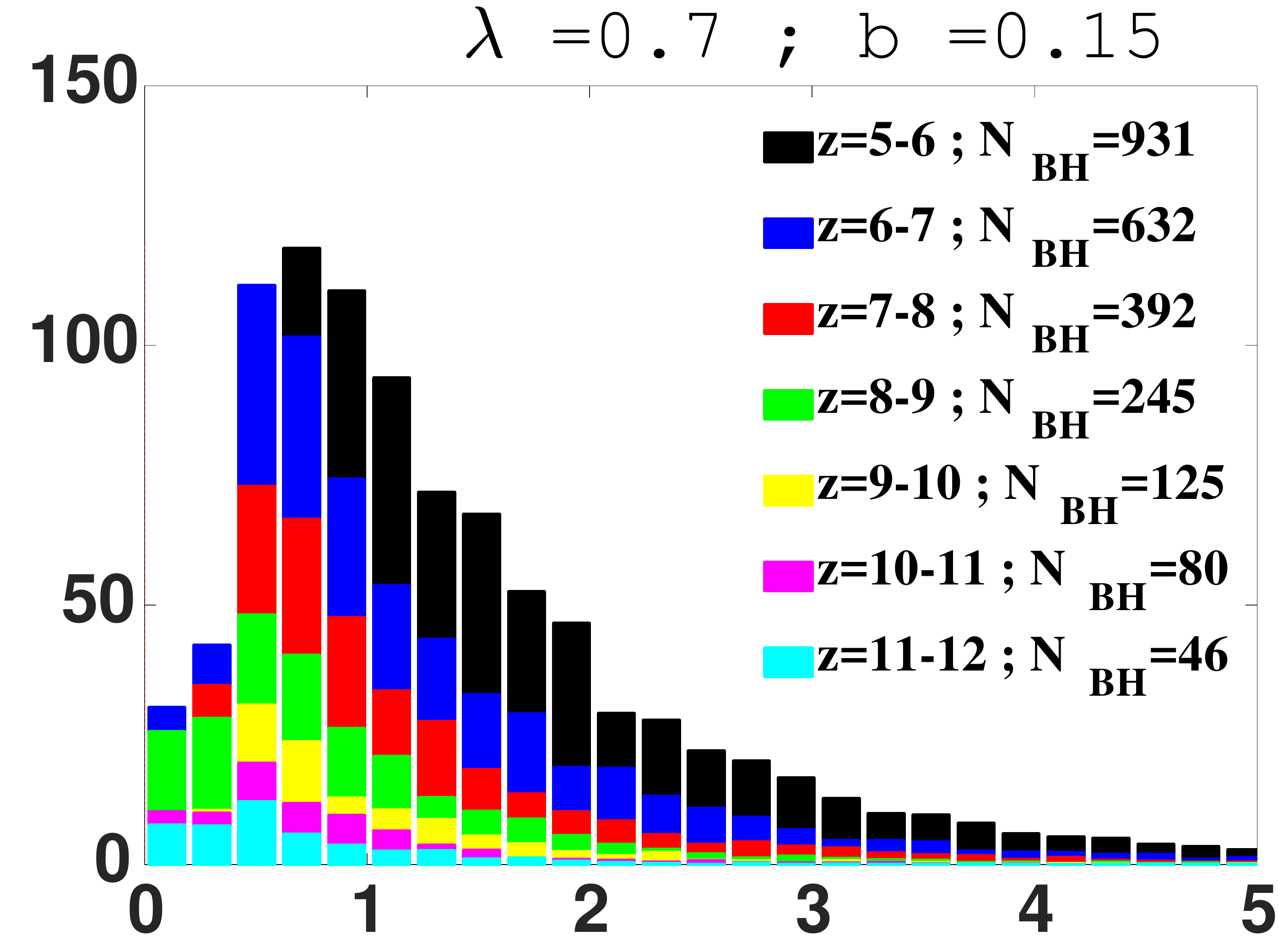}} 
{\includegraphics[width=0.22\columnwidth]{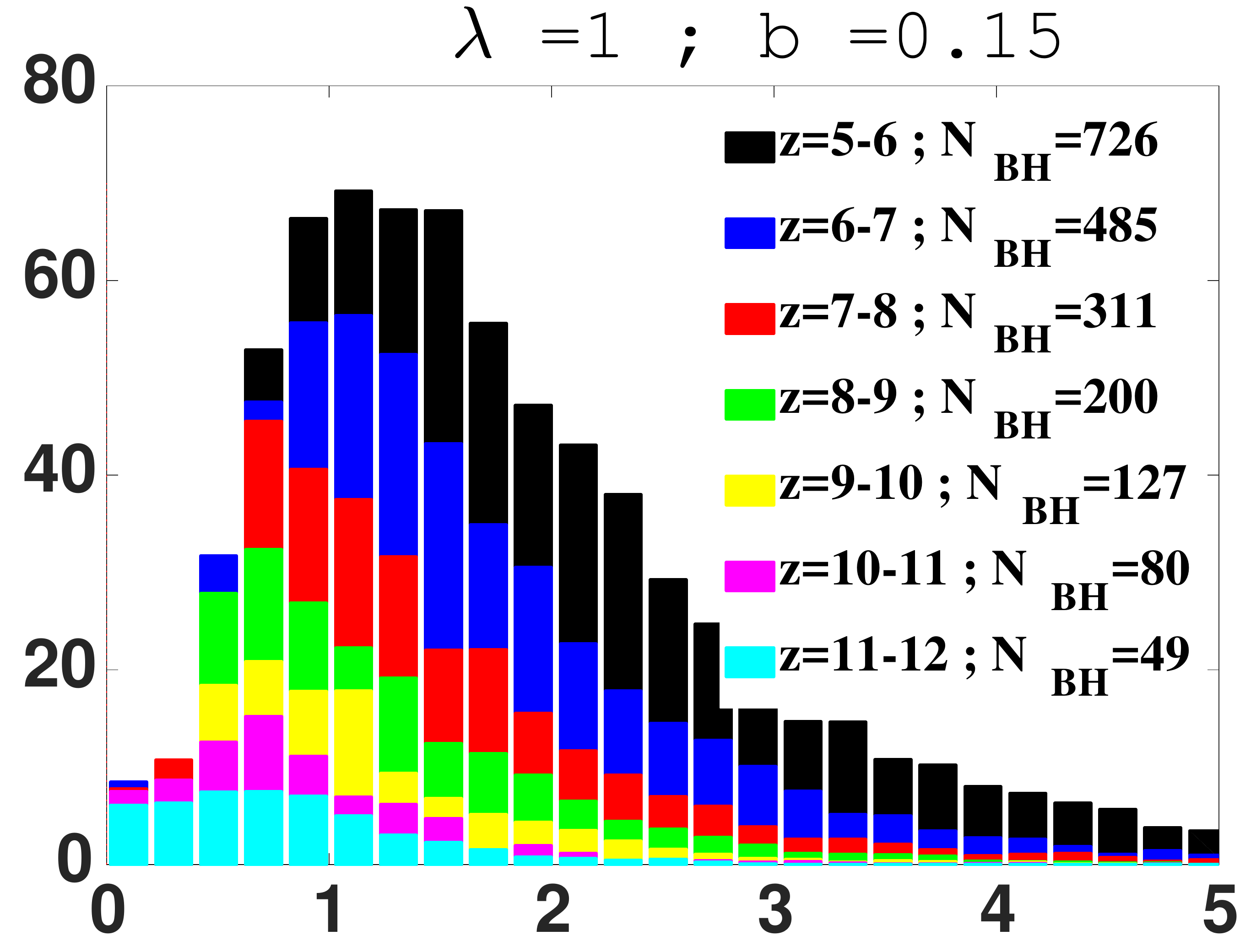}} \\
{\includegraphics[width=0.22\columnwidth]{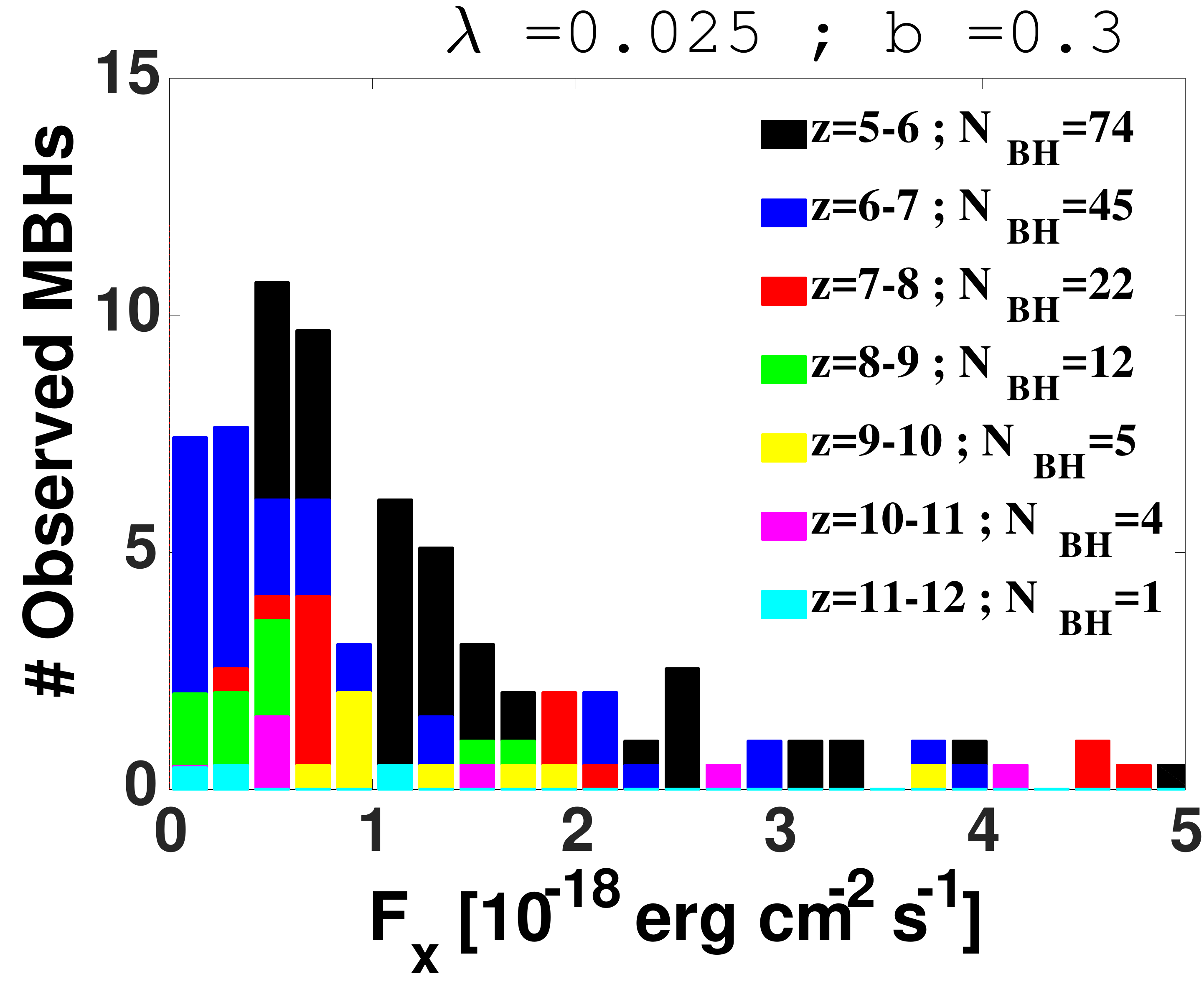}} 
{\includegraphics[width=0.22\columnwidth]{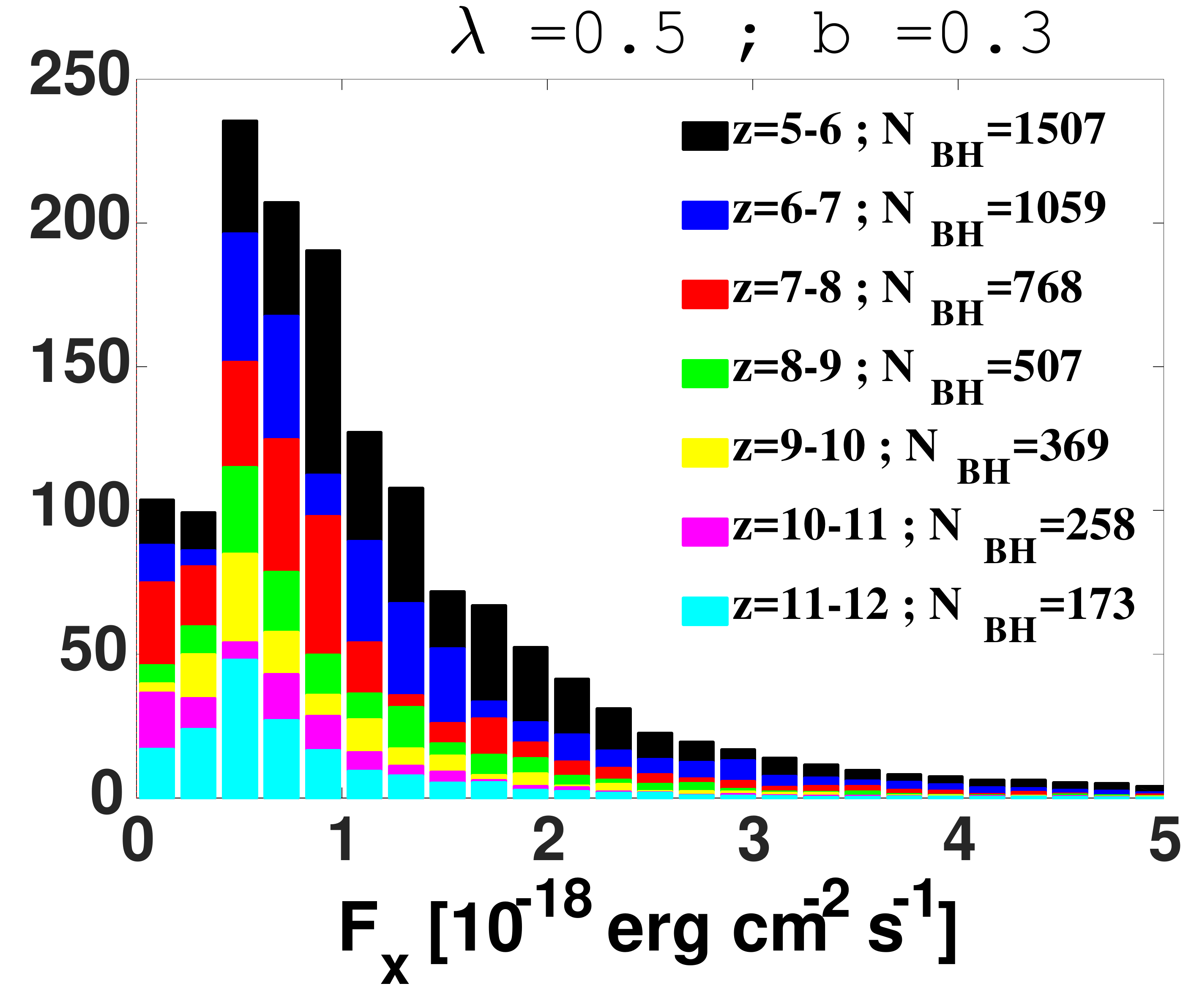}} 
{\includegraphics[width=0.22\columnwidth]{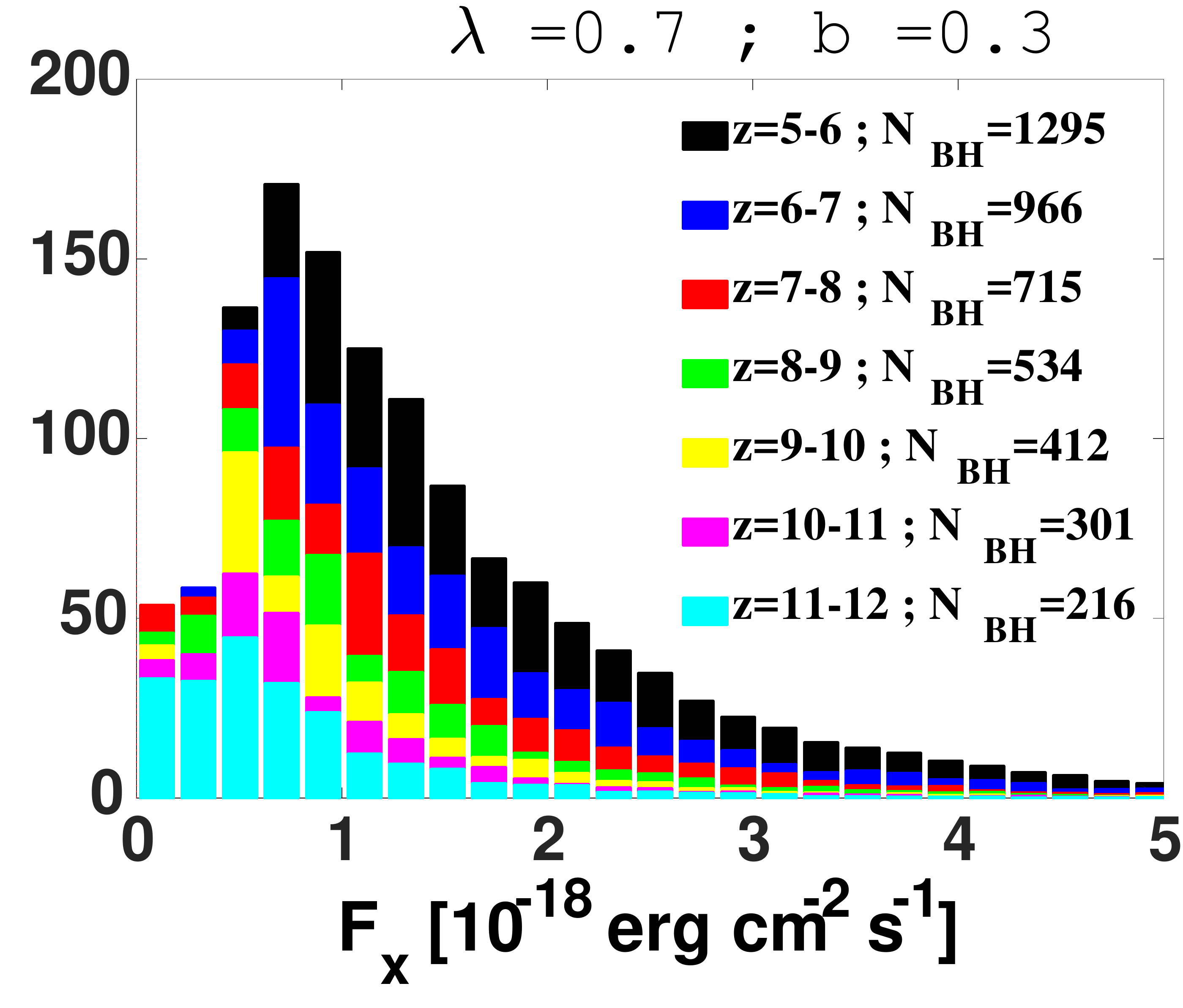}} 
{\includegraphics[width=0.22\columnwidth]{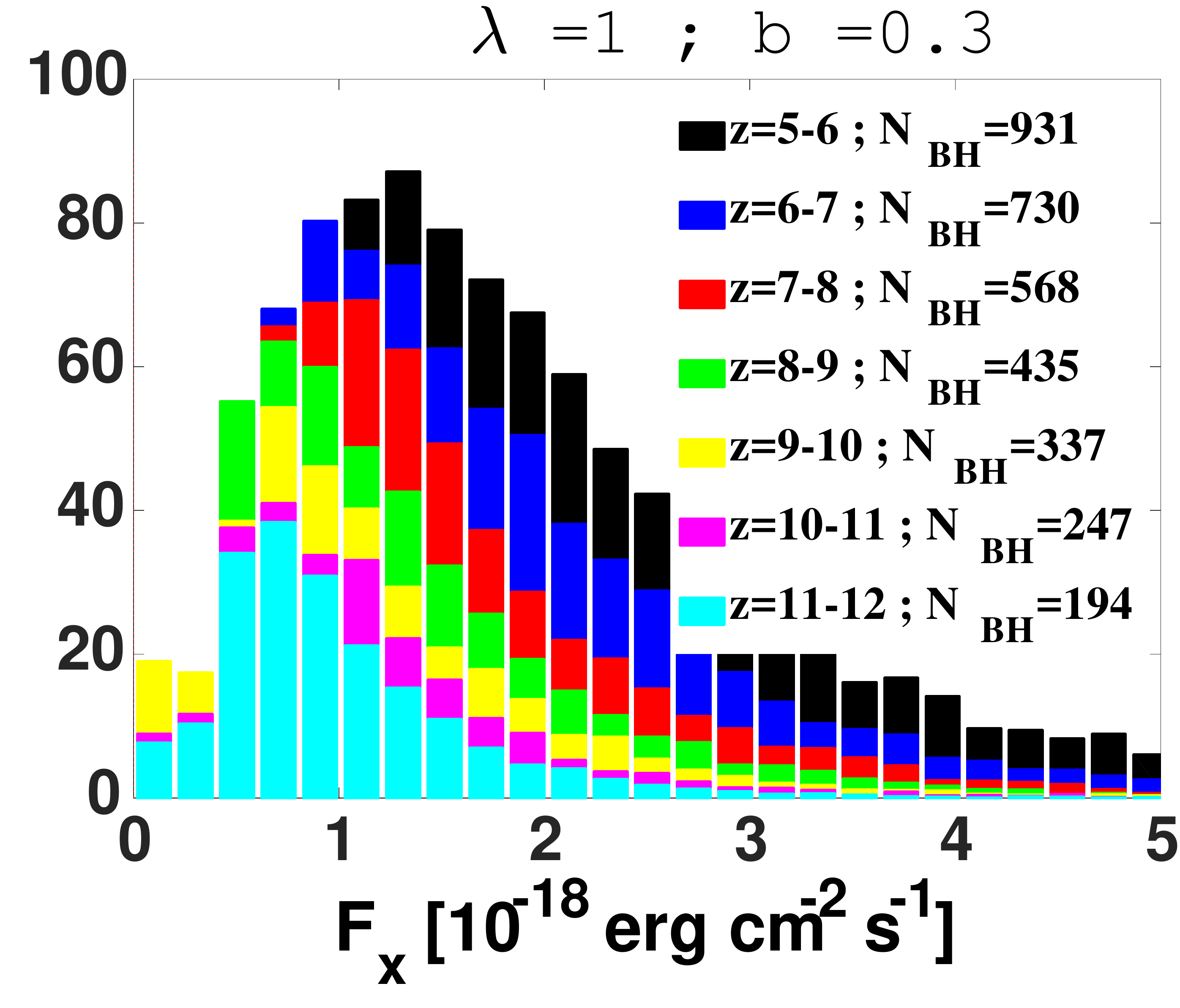}} \\
\end{center}
\caption{Expected X-ray flux distributions for heavy seed models as observed with Lynx.}
\label{fig:HeavyFxDist}
\end{figure}

In the previous section we used physically motivated models and relations to estimate the number of observable MBHs and their flux distribution as a function of redshift. Our models are represented by two parameters: the mass scaling parameter \textit{X}$=\frac{\mathrm{M}_{BH}^z}{\mathrm{M}_{BH}^{z=0}}$, and the luminosity scaling parameter $\lambda=\frac{L}{L_{edd}}$. In some of the cases, the parameters we use for each model are drawn from a distribution and will vary for different objects.\\
The variance and uncertainty in the parameters we use blur the differences in the observables between models. We take this effect into account by performing a MC simulation for each model. The results of the MC run allow us to assess the distribution and variance in the flux-redshift phase space in an attempt to estimate whether current or future observations allow us to distinguish between seeding and growth models. 
For each model, we run a series of MC realizations with the following assumptions:
\begin{enumerate}
\item The number of observable MBHs in the FoV follows a Poisson distribution, with $\overline{N}=\eta_{oc}\times \overline{N^{\dagger}_{obs}(z)}$.\\
$\overline{N^{\dagger}_{obs}(z)}$ is the average number of observable MBHs when varying $\eta$ between a value of 2 and 4, and $\eta_{oc}$ is the occupancy factor. \cite{Volonteri2008} argue that the host occupancy is of the order of $\sim10\%$ ($4\%-25\%$ in models presented), with the occupancy increasing as a function of host mass. We therefore assume $\eta_{oc}$ follows a normal distribution with $\overline\eta_{oc}=0.14$ and $\sigma=0.04$. 
\item BH masses follow a normal distribution with a mean equals to the weighted average BH mass, and $\sigma$ being the standard deviation of the distribution weighted by the cumulative halo density function.
\item The luminosity scaling parameter $\lambda=\frac{L}{L_{edd}}$ follows a log-normal distribution with $\nu=0.4$ \citep{Kelly2010}. We apply an upper limit of $\lambda=1$ to avoid cases of super-Eddington accretion (see section 4 for an analysis of models allowing super-Eddington luminosities).
\item To estimate the flux attenuation due to circum-nuclear and gas absorption, we assume a normal distribution for column densities with an average of N$_H=10^{23}$cm$^{-2}$ , and $\sigma=5\times10^{22}$cm$^{-2}$ \citep{Vito2013,Vito2014}. We model the amount  of signal attenuation using the \textit{xspec} model \textit{plcabs}\footnote{https://heasarc.gsfc.nasa.gov/xanadu/xspec/manual/XSmodelPlcabs.html}, which assumes a power-law SED emission transmitted through a spherical distribution of cold and dense matter, taking into account Compton scattering \citep{Yaqoob1997,Vito2013}.
\item In order to estimate the duty cycle for each MBH realized in the run, we calculate $v_{circ}^{max}$ of the MBH host, and evolve it to a redshift $z=2$ according to results from the Bolshoi N-body simulation (see section 2.2). We calculate M$_{BH}$ for the evolved halo assuming the M$_{BH}-\sigma$ relation has settled to the one observed in the local universe (\textit{i.e.,} \textit{X}$=1$). We derive the amount of time $t_{acc}$ the MBH will need to accrete the mass difference, assuming the MBH specific value of $\lambda$ and accretion efficiency $\epsilon=0.2$ \citep[\textit{i.e.}, coherent accretion,][]{Berti2008}. We assume contribution from MBH mergers is negligible \citep[\textit{e.g.,}][]{Volonteri2009,Kulier2015}. The duty cycle is than estimated according to $\eta_{dc}=\frac{t_{acc}}{\Delta t}$, $\Delta t$ being the time elapsed between the redshift at which the MBH is observed, and redshift $z=2$.  
\end{enumerate}
For each model, characterised by $\lambda$ and the mass scaling parameter \textit{X}, we run 500 MC realizations. The output of our MC run is a list of MBHs, each characterized by mass, redshift, observed X-ray flux, and duty cycle. Since we study MBHs past the seeding stage, and only after initial host-MBH co-evolution, we verify for each BH whether its mass is above three times the seeding mechanism minimum mass, which we assume to be $5\times10^4\mathrm{M}_{\odot}$ for the heavy model \citep{Haehnelt1993, Loeb1994, Eisenstein1995,Bromm2003} and $100\mathrm{M}_{\odot}$ for the light seed model \citep{Bromm1999,Heger2003,Gao2007}. We also verify whether the observed flux from the object is indeed above the detection limit of the experiment we investigate, after we include circum-nuclear absorption into account. Conservatively, we discard a simulated MBH in case it did not pass the aforementioned criteria.\\ 
With the MC output in hand, we count the number of observable MBHs for each telescope, taking into account the duty cycle of each source. The results are shown in Fig. \ref{fig:N_bh}. For Lynx, the high number of observed objects allows us examine the expected X-ray flux distribution. We therefore bin the data in redshift (7 bins) and flux (50 bins)\footnote{The number of bins is chosen to optimize the statistical significance of our results as a function of phase space resolution. When deriving the expected number of observed objects, as well as the flux distributions, we average over the 500 MC realizations with the statistical error being the standard deviation of the 500 MC runs.} to derive the redshift and model dependent X-ray flux distributions, see Figs. \ref{fig:LightFxDist} - \ref{fig:HeavyFxDist}. 


 \begin{deluxetable*}{cccc}
 \tablehead{\colhead {Parameter} & \colhead {Value} & \colhead {Distribution} & \colhead {Comments} }
 \startdata
$\lambda=\frac{L}{L_{edd}}$ & $0.1-1$ & Log-Normal & -  \\
X$=\frac{\mathrm{M}_{BH}^z}{\mathrm{M}_{BH}^{z=0}}$ & $b=-0.3 $-$ 0.3$ & - & $X=e^{b(z-2)}$  \\
$\eta=\frac{v_{circ}^{max}}{\sigma}$ & $2-4$ & - & \tablenotemark{a}\\
$\eta_{dc}$ &   &  & Growth into local M$_{BH}-\sigma$ relation \\
$\eta_{oc}$ & $14\%$ & Normal & $\sigma=4\%$ \tablenotemark{b}   \\
Column Density & N$_H=10^{23}$cm$^{-2}$ & Normal & $\sigma=5\times10^{22}$cm$^{-2}$ \tablenotemark{c}\ \\
\hline
$\lambda=\frac{L}{L_{edd}}$ & $1.7-2.1$ & Log-Normal & Slim Disk Models \\
\hline
$\lambda=\frac{L}{L_{edd}}$ &  $\overline{\lambda}=-2.49+1.93\log(1+z)$ & Log-Normal & Torque Limited  Models \tablenotemark{d}\\
 \enddata
\label{tab:Parameters}
\tablenotetext{a}{\cite{Padmanabhan2004,Dutton2004,Kravtsov2009}}
\tablenotetext{b}{\cite{Volonteri2008}}
\tablenotetext{c}{\cite{Yaqoob1997,Vito2013}}
\tablenotetext{d}{\cite{Angles2013,Angles2015}}
\end{deluxetable*}

\section{Analysis and Discussion}
\begin{figure}[]
\begin{center}																  
{\includegraphics[width=0.4\columnwidth]{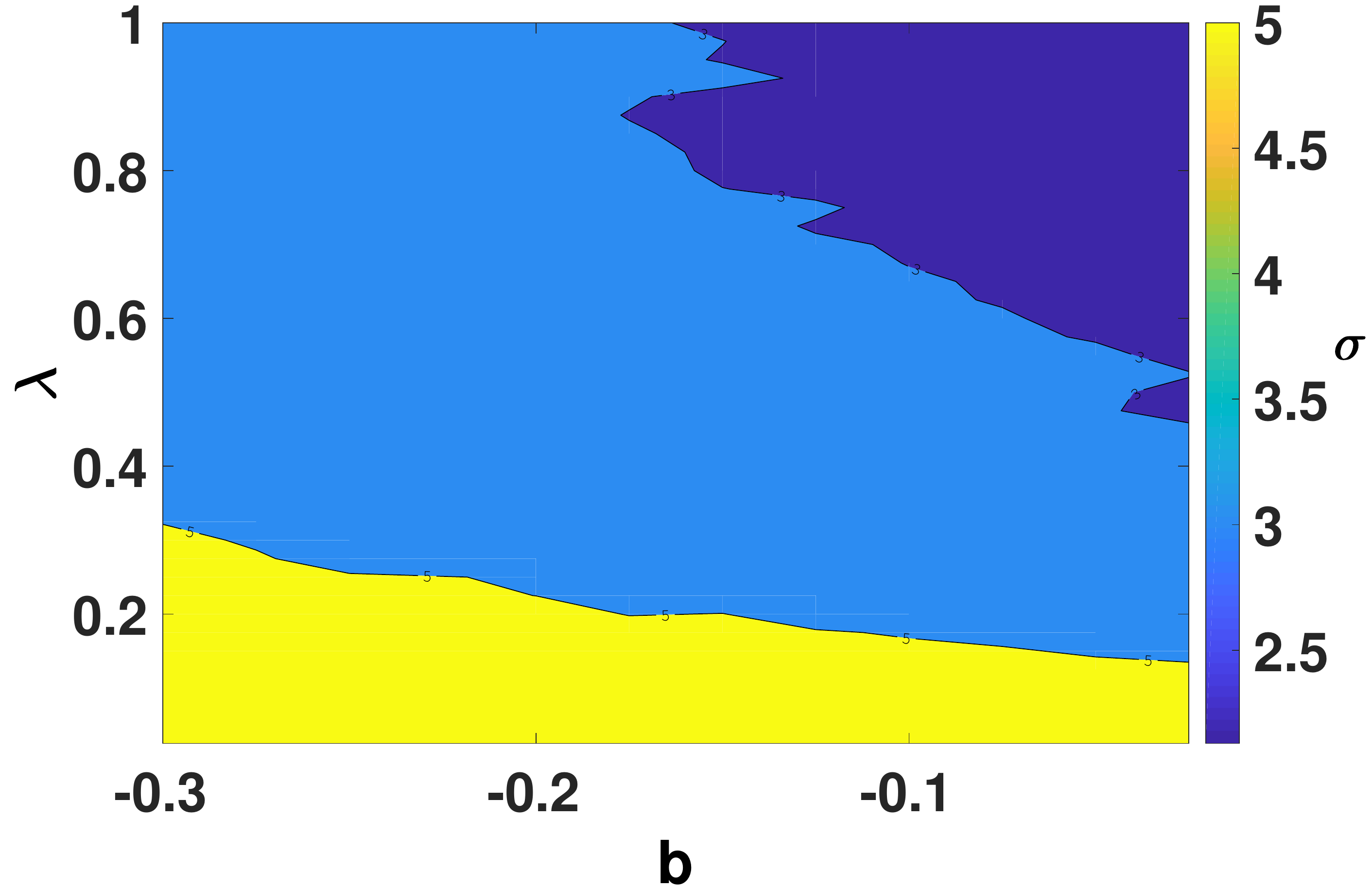}} 
{\includegraphics[width=0.4\columnwidth]{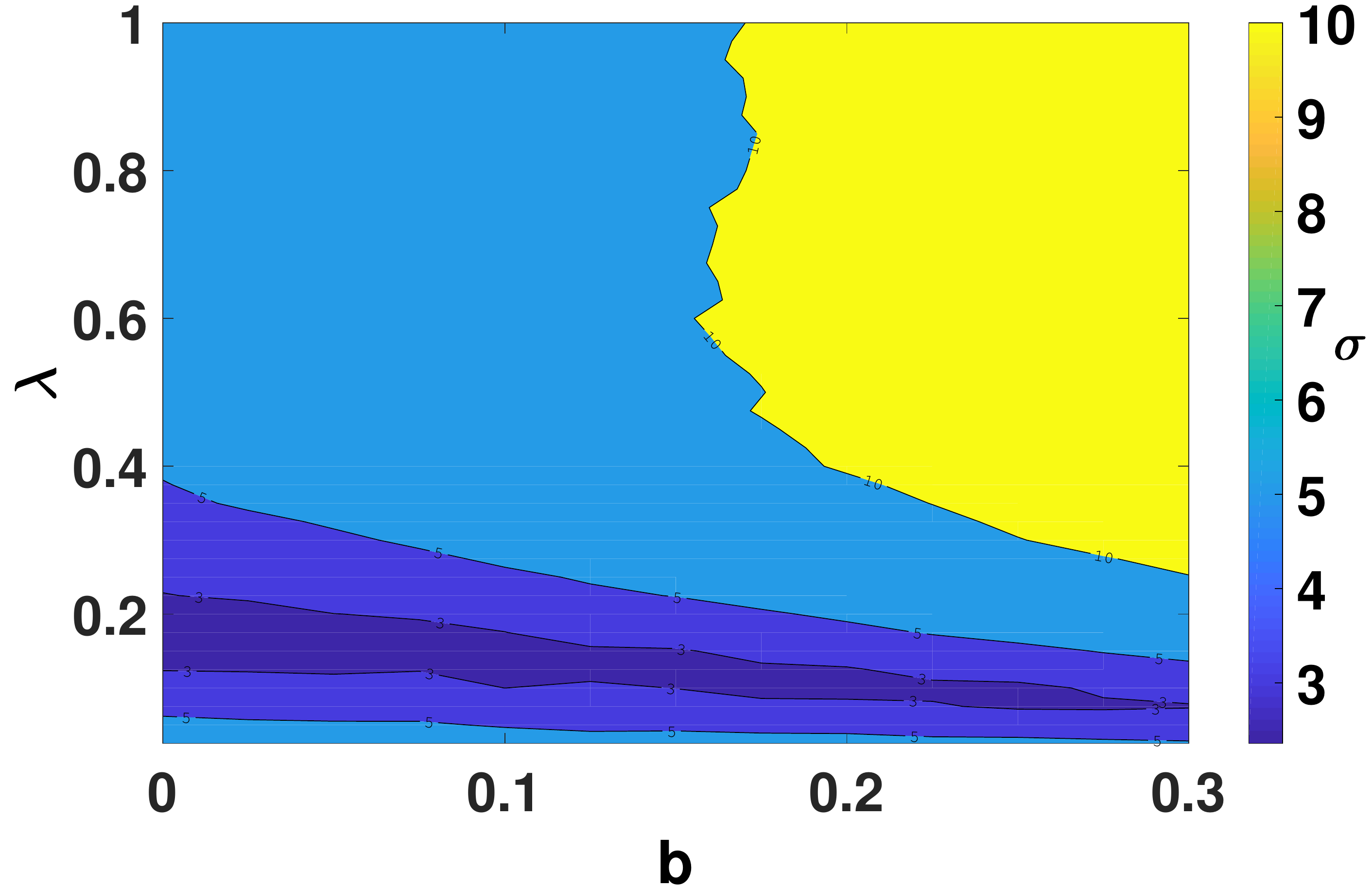}} \\
{\includegraphics[width=0.4\columnwidth]{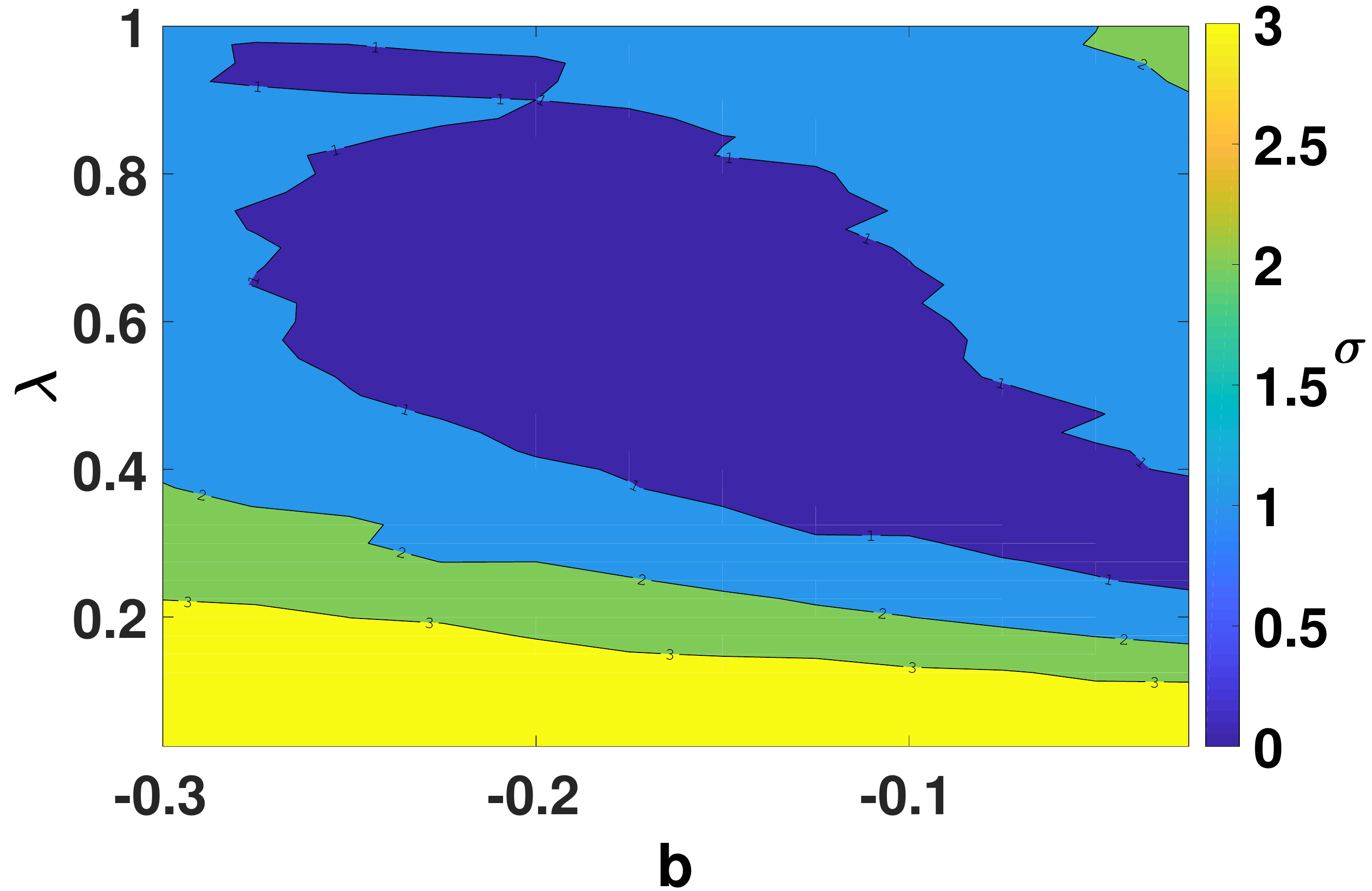}} 
{\includegraphics[width=0.4\columnwidth]{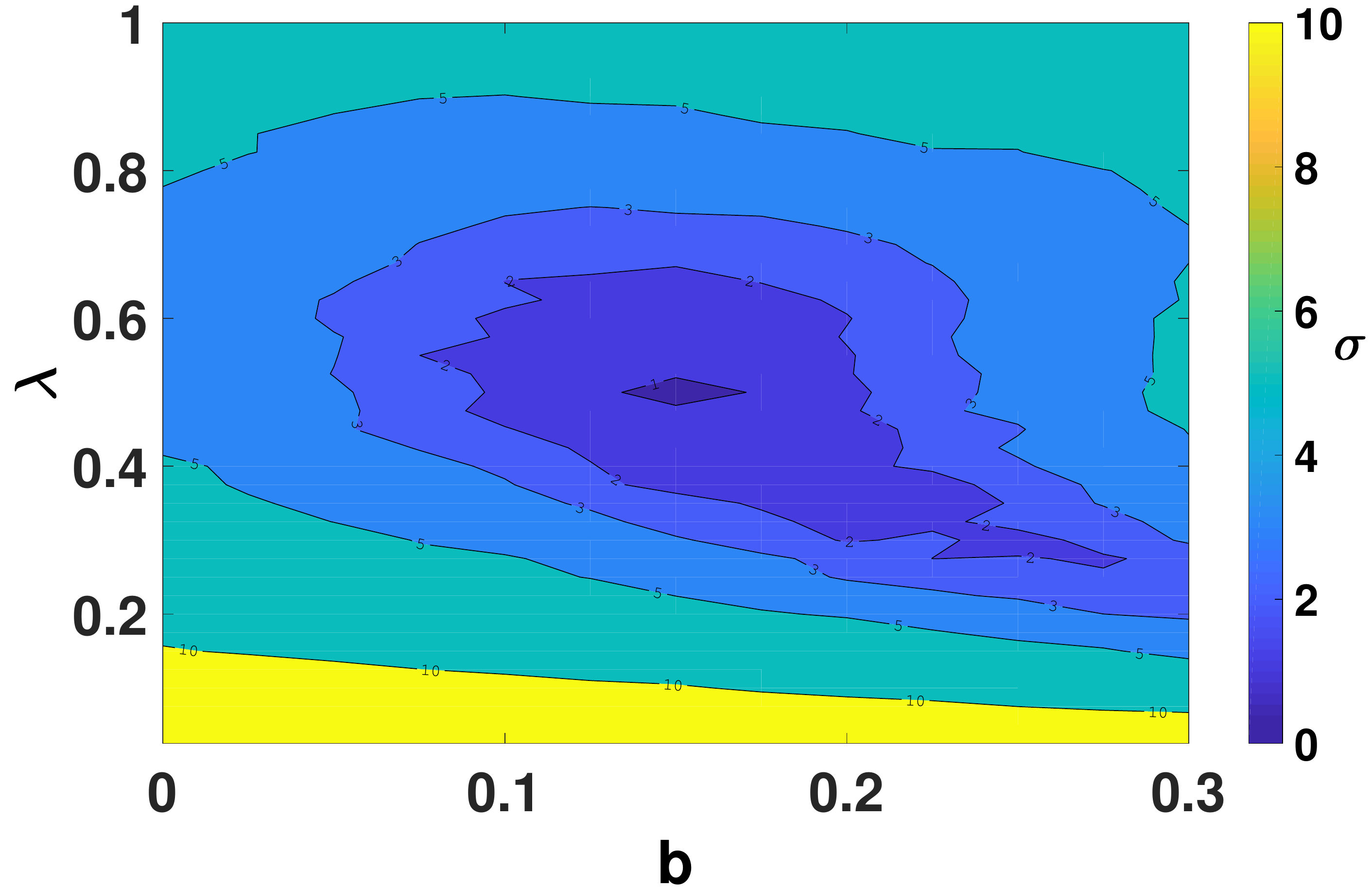}} \\
\end{center}
\caption{Statistical discrimination based on X-ray flux distributions. We perform a $\chi^2$ test for each family of models. Top left: Light seed models with a heavy seed model of [$b=0.05$ ; $\lambda=0.2$] as the null hypothesis. Top right: Heavy seed models with the light seed model of [$b=-0.05$ ; $\lambda=0.8$] as the null hypothesis. A stististical discrimination between heavy and light seeds models at the $3\sigma$ level is achieved across most of teh phase space investigated in this work.
Bottom left: Light seed models with [$b=-0.15 ; \lambda=0.5$] as the null hypothesis.  Bottom Right: Heavy seed models with [$b=0.15 ; \lambda=0.5$] as the null hypothesis. Discrimination within a family of models can direct us to the phase space in which plausible scenarios reside.}
\label{fig:Chi2_surveyor}
\end{figure}

The expected number of detected objects, and the redshift dependent X-ray flux distributions vary significantly between the two telescopes, and between models. We analyze the MC simulation results in the context of our capability to differentiate between the various models.\\ 
For all simulated models, the number of MBHs expected to be detected by the Chandra X-ray observatory is few at most. Only models in which the MBH luminosities are comparable to the Eddington limit yield any detectable objects. Due to the small numbers of observable sources, there is no significant statistical difference between the light and heavy seed models, as shown by the top panels of Fig. \ref{fig:N_bh}. Indeed, analysis of CDFS data \citep[\textit{e.g.,}][]{Vito2013,Treister2013,Weigel2015} show no clear detection of intermediate MBHs in the FoV in agreement with our simulation results. Our findings further emphasize the limitations of current observations, owing to the small effective area of the telescope and the small FoV.\\
Considering instead Lynx detection limit and FoV, the number of detectable sources varies from several dozens (light seed models with $\lambda\sim0.025$) to $\sim4000$  (heavy seed models with $\lambda\sim0.7$), as illustrated in the bottom panels of Fig. \ref{fig:N_bh}. The high number of detectable MBHs allows to examine the X-ray flux distribution at various redshifts, as shown in Figs. \ref{fig:LightFxDist}-\ref{fig:HeavyFxDist}. The results exhibit three dominant trends: 
(\textit{i}) in the light seed models, the peak of the distribution is around the detection limit; this seems to be in agreement with the underlying halo density, as halos of lower $v_{circ}^{max}$ are more abundant than halos of higher $v_{circ}^{max}$, and the peak is determined by the limiting flux of the experiment under study; in the heavy seed models, the peak of the flux distribution moves to higher flux values for higher values of $\lambda$ in all models, as most of the MBHs have evolved above the mass associated with the flux limit, while the smaller halos, which in general would have hosted a detectable MBH, do not host such an object;  
(\textit{ii}) the maximum number of objects is detected in models with $\lambda=0.6-0.7$; whereas in general one would assume that the maximum number of objects would be detectable in models in which the MBHs shine at their Eddington luminosity, these models have a lower average duty cycle and therefore at a certain value of $\lambda$ the low duty cycle overcomes the higher accretion rate; this effect is more significant for higher values of the mass scaling parameter \textit{X}, since the median duty cycle approaches an asymptotic value of unity for smaller values of \textit{X};
(\textit{iii}) for higher values of the mass scaling parameter, the number of objects detectable at very high redshifts ($10<z<12$) increases from zero ($b=0, \lambda=0.025$) to $\sim200$ objects ($b=0.3, \lambda=0.7-1$), while in the case of light seed models, no significant number of objects are expected to be detected above redshift $z\sim10$; this is is agreement with published models, arguing that detection of Pop-III SMBH seeds at high redshifts is unlikely \citep{Natarajan2017}. We emphasize that there is a clear degeneracy between $\lambda$ and $b$ when examining the overall number of detected objects, but the relative number of observed objects at a given redshift is a good indicator for the evolution and seeding mechanism of MBHs.\\
As can be seen from the redshift dependent X-ray flux distributions, a clear difference between the simulated light and heavy seed models is observed in most cases. In order to get a quantitative understanding of the discrimination power in some limiting cases, we perform $\chi^2$ tests for the light and heavy seed models. For the light seed models we assume the null hypothesis is the heavy seed model [$b=0.05 ; \lambda=0.2$] while for the heavy seed models we assume the null hypothesis is the light seed model [$b=-0.05 ; \lambda=0.8$]. We choose these two null hypotheses as the flux distributions derived for these models are similar to the expected distribution from the opposite group of models, but do not represent extreme cases.
The results, shown the top panels of Fig. \ref{fig:Chi2_surveyor} imply clearly that a discrimination at a level of $3\sigma$ can be achieved across most of the phase space, excluding some extreme cases for each group of models. 
In order to quantify our ability to constrain the parameters used to characterize a specific model, we perform $\chi^2$ tests within a given family of models. For the light seed models, we assume a null hypothesis of [$b=-0.15$ ; $\lambda=0.5$], while for the heavy seed models the null hypothesis is [$b=0.15$ ; $\lambda=0.5$], see the bottom panels of Fig. \ref{fig:Chi2_surveyor}. 
We find that while the simulation is clearly degenerate for several parameters (mass scale parameter, luminosity scale parameter, seed generation efficiency and duty cycle), the observational results from Lynx can direct us to the phase space in which plausible scenarios reside within a specific family of models.

\subsection{Additional constraints from the cosmic X-ray background}
The observed X-ray background (XRB) is thought to be the result of two main contributions: (\textit{i}) the galactic X-ray background emitted by hot gas in the solar neighborhood governs the soft end of the XRB ($<0.3\,$keV); and (\textit{ii}) the Cosmic X-ray background (CXRB) associated with the accumulated emission from unresolved sources, mainly obscured and unobscured  active galactic nuclei (AGNs), normal galaxies, and the IGM \citep{Cappelluti2012}.  Current Chandra measurements of the CXRB yield a value of $4.6\pm0.3\times10^{-12}\,\mathrm{erg\,cm^{-2}\,s^{-1}\,deg^{-2}}$ from which $23\pm3\%$ is unresolved \citep[$1-2$\,keV band;][]{Hickox2006,Hickox2007}. Several studies indicate that at low flux levels the AGNs contribution to the CXRB drops, whereas the contribution from normal galaxies rises. At the CDFS flux limit $46\%\pm5\%$ of the source counts are normal galaxies \citep{Lehmer2012}. We therefore examine next whether any of the models in our MC simulation violates the CXRB limit, and whether current or future observations can shed some light on the origins of the CXRB.

We ran MC simulations for all models, with an upper limit set to the flux limit in CDFS observations, as MBHs that in general can be detected as point sources by Chandra do not contribute to the unresolved CXRB. We do not apply a lower limit on the point source flux but keep the mass lower limit associated with each seeding model, as this is associated with the physics of SMBH seed generation and not with any observational limit.\\
Our results show that in all cases, the CXRB contribution from AGNs at high redshifts lies below the current upper limit. For the light seed models we find the MBH contribution varies from $10\%\pm5\%$ for the [$b=-0.3$ ; $\lambda=0.025$] model to $24\%\pm9\%$ for the [$b=0$ ; $\lambda=1$] model, while for the heavy seed models we find contributions at the level of $18\%\pm6\%$ for the [$b=0$ ; $\lambda=0.025$]  model, and $81\%\pm18\%$ for the [$b=0.3$ ; $\lambda=0.575$] model. We emphasize that when considering the diminishing contributions from AGNs at the Chandra low flux limit, heavy seed models with $\lambda=0.3-0.8$ and $b>0.225$, provide a factor of $1.5-2$ more flux than expected (for higher values of $\lambda$, the duty cycle drops and so the overall integrated flux drops as well, see previous section). 
We also investigate what is the expected contribution from unresolved AGNs to the unresolved CXRB assuming we perform observations down to Lynx flux limit. We ran the same MC simulation with an upper limit set to $1\times10^{-19}\,\mathrm{erg\,cm^{-2}\,s^{-1}}$. The results indicate that for the heavy seed models the contribution from unresolved AGNs will be below $5\%$ of the total unresolved CXRB, and so most of the AGNs contribution falls between Lynx and Chandra detection limits. For the light seed models we find that a significant fraction of the MBHs contributing to the unresolved CXRB, will remain below the detection limit, in agreement with \cite{Natarajan2017}. We conclude that estimates of the contribution of MBHs at high redshift to the CXRB, combined with deep observations with a future high resolution X-ray telescope, can yield additional constraints on the scenarios by which SMBH seeds are generated and evolve.

\section{Other models}
\subsection{Super-Eddington Accretion}
\begin{figure}[]
\centering
{\includegraphics[width=0.45\columnwidth]{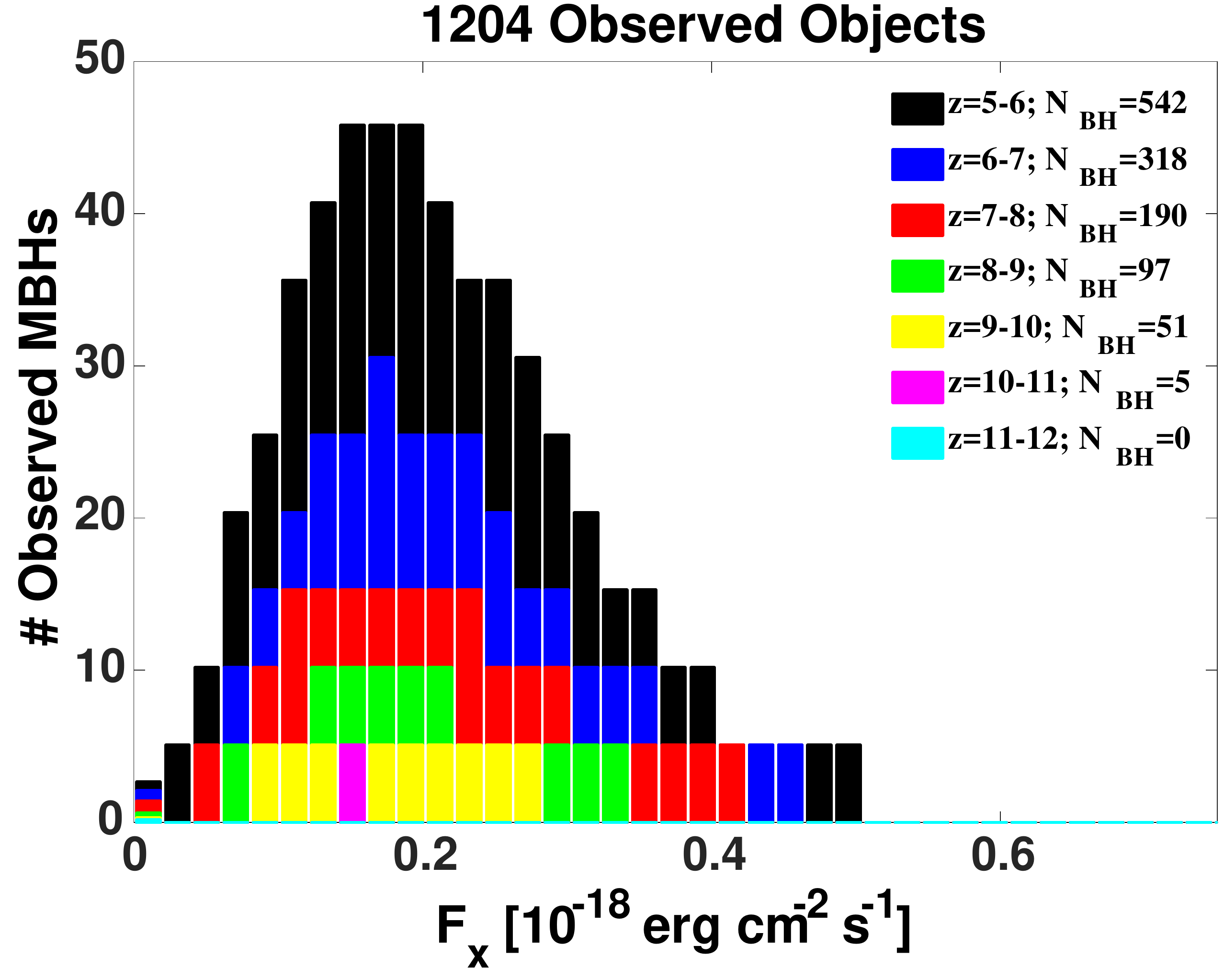}} 
\caption{The expected X-ray flux distributions for observations with Lynx for slim disk models. The models investigated are equivalent to a [$b=0$ ; $\lambda=1.7-2.1$] in terms of the parameters presented in sections 2 and 3. }
\label{fig:fxDist_SuperEd}
\end{figure}

While the scenario in which a SMBH seed is the outcome of the collapse of a gas cloud to a $\sim10^5$M$_{\odot}$ BH is a possible solution to the problem raised by the discovery of quasars at $z\sim6-7$, an alternative is the growth of SMBH seeds through super-Eddington accretion in several short episodes \citep[$\sim\mathrm{O}(10)\,$Myrs, \textit{e.g.,}][]{Wyithe2012,Tanaka2014,Castell2016}. We turn now to investigate a family of models which assumes mild super-Eddington accretion by a SMBH seed, namely the slim disk models \citep{Abramowicz1988,Jiang2014,Jiang2017}.  \cite{Madau2014} analyze the case of a slim disk solution, in which the viscosity generated heat in the accretion disk is not radiated away immediately, but is advected into the central SMBH seed. In this model, mild super-Eddington accretion can take place for several Myrs, thus shortening the characteristic e-folding time of the MBH. \cite{Madau2014} analyze two cases in which a Pop-III seed\footnote{While slim disk models were analyzed in the literature in the context of both heavy and light seed models \citep{Natarajan2017}, we follow \cite{Madau2014} and focus on the application of slim disk accretion in the light seed scenarios.} will grow into a quasar by $z\sim7$: (\textit{i}) three major episodes each lasting $50\,$Myr in which $\lambda=3\,$ followed by a $100\,$Myr period of quiescence (duty cycle = $0.5\,$); and (\textit{ii}) five major accretion episodes of $20\,$Myr in which $\lambda=4$, followed by $100\,$Myr period of quiescence (duty cycle = $0.2\,$). 
In an attempt to examine observables for these scenarios, we repeat the MC simulation described earlier. We assume the M$_{BH}-\sigma$ relation at high redshift is similar on average to the one observed in the local universe (\textit{i.e.,} $b=0$ in terms of the parameters used for the sub-Eddington models), as MBHs at the end of an accretion phase will be more massive with respect to their host halo, while MBHs at the end of a quiescent phase will be less massive with respect to their host halo.
While the two cases discussed above are able to produce quasars by $z\sim7$, the relative comoving number density of quasars to MBHs at $z>6$ is expected to be $10^{-6}-10^{-5}$ \citep[\textit{e.g.,}][]{Bromley2004}, and so we choose a combination of $\lambda$ and duty cycle that generates the expected density ratio. A bootstrap analysis of models in which $\lambda$ has a log-normal distribution, the duty cycle is constrained so that the M$_{BH}-\sigma$ relation observed in the local universe holds for all redshifts, and the two models analyzed by \cite{Madau2014} yield the expected density of quasars at $z\sim6$, constrains us to a phase space in which $\lambda=1.7-2.1$. All other assumptions in our MC simulation are not changed (see section 2.4). We perform $500$ MC runs, and average over the results, shown in Fig. \ref{fig:fxDist_SuperEd}. The overall number of sources expected to be detected is similar to the sub-Eddington heavy seed models. The flux-redshift distribution, however, is different, with the peak at a higher value relative to the sub-Eddington cases. The lower skewness of distribution is due to our assumption on the probability distribution of $\lambda$, which is more symmetric for higher values of $\lambda$. Further analysis needs to be performed in order to assess the significance of our result, if one assumes a different PDF for $\lambda$. In addition, we find more sources to be observable at high redshifts ($z>7$) than in the sub-Eddington case, since in general the sources are expected to be brighter due to the high value of the luminosity scaling parameter, and not due to the higher mass of the SMBH seeds as in the heavy seed models (with the degeneracy between various parameters in the models evident again in this case). We repeat the $\chi^2$ test with the null hypothesis the slim disk models X-ray flux distributions. Our goal is to get a quantitative understanding of the difference between the heavy seeds sub-Eddington and slim disk models. The results suggests that a discrimination at the level of $>5\sigma$ can be achieved across all of the phase space we examine in section 3.\\ 
Summing the contribution from AGNs below the Chandra flux limit, we find that for the slim disk models, SMBH seeds will contribute $28\%\pm{6\%}$ of the unresolved CXRB - well below current upper limits from CDFS \citep{Lehmer2012}, and similar to the results derived for the sub-Eddington light seed models described in section 3.1\,. 
\subsection{Torque limited accretion models}
\begin{figure}[]
\begin{center}
{\includegraphics[width=0.45\columnwidth]{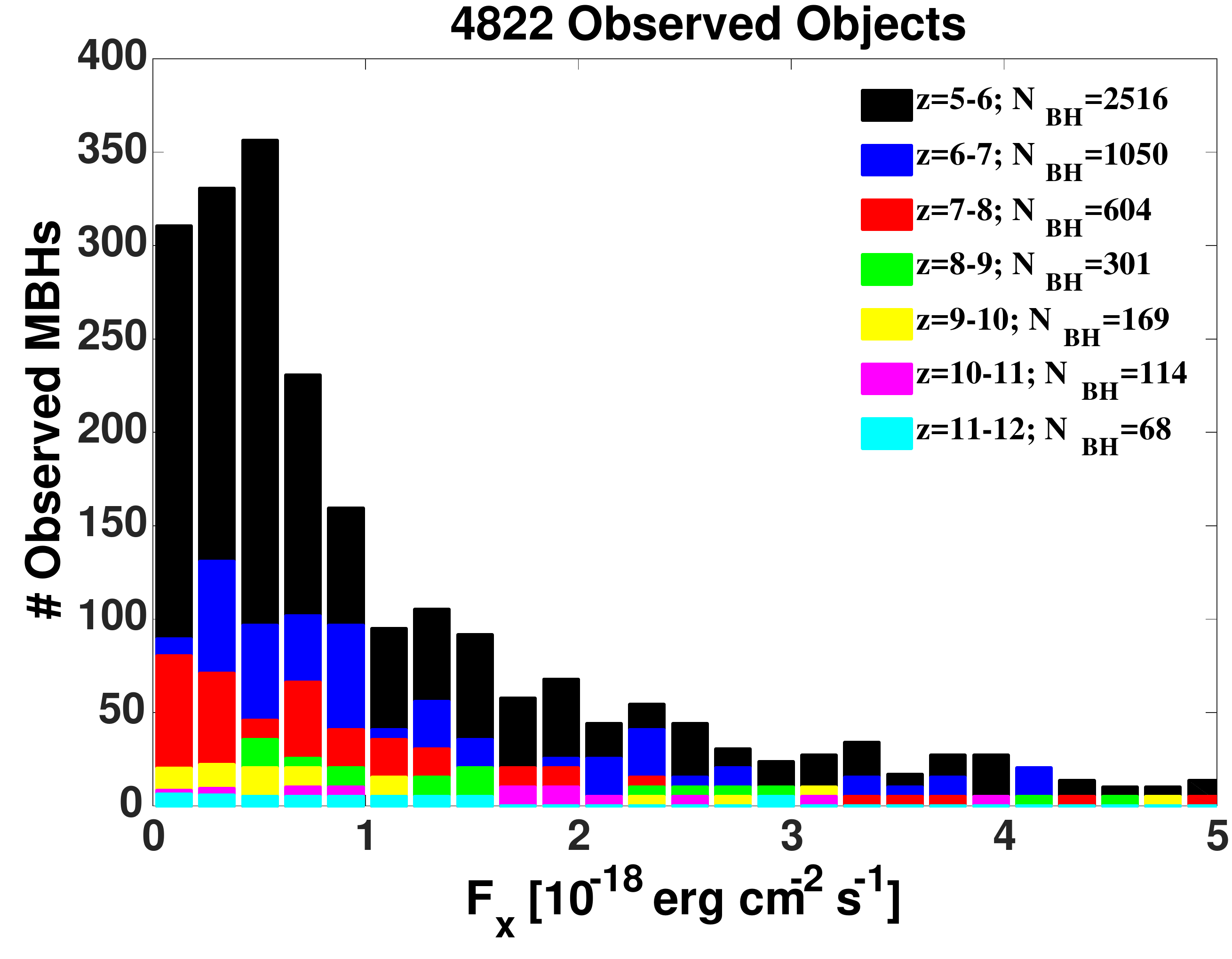}}\\
\caption{The expected X-ray flux distributions for observations taken with Lynx for torque limited growth of heavy seeds anchored at z=12.}
\label{fig:fxDist_torqueLimited}
\end{center}
\end{figure}

The observed scaling relations between SMBHs and their host galaxies are often explained by self regulated growth of the MBH seeds and co-evolution with the hosts.
 Some of the models induce self regulation by strongly coupling winds from the accreting MBHs to the gas supply in the galactic scale, \citep[\textit{e.g.,}][]{Wyithe2003,Springel2005,Debuhr2011,Hambrick2011,Kim2011,Dubois2012,Choi2012}. However, there is no direct evidence for the self regulation to be caused by feedback from the accretion process. In the SMBH growth models studies by \cite{Angles2013,Angles2015}, no direct feedback from the MBH accretion process is assumed. Instead, galaxy scale gravitational torques induced by interactions or self gravitating disks are the limiting factors in supplying gas from the galactic scale to the MBH sphere of influence, thus limiting the MBH growth. These models yield MBHs that evolve naturally onto the observed galaxy - SMBH scaling relations, regardless of the initial conditions. While \cite{Angles2015} shows a detailed analysis at lower redshifts than the ones we examine in this work, we extend their results to higher redshifts using the methdology described in section 2, with the following changes:
\begin{enumerate}
\item We anchor the MBH to $v_{circ}^{max}$ at $z=12$, assuming $\eta=2-4$. The MBHs are ten times heavier (lighter) than inferred from the M$_{BH}-\sigma$ relation observed at $z=0$. This represents heavy (light) seeds.
\item Heavy seeds grow into the M$_{BH}-\sigma$ relation in $1.5\times t_{Hubble}$, where $t_{Hubble}=0.96\mathrm{Gyr}[(1+z)/z]^{-3/2}$ is the Hubble time at redshift $z$, while light seeds grow into the local M$_{BH}-\sigma$ relation in $0.5\times t_{Hubble}$. \cite{Angles2015} show that this assumption is independent on the redshift at which we anchor the SMBH seeds to the M$_{BH}-\sigma$ relation.
\item We examine only MBH population during initial co-evolution until the local scaling relations have been established. For the heavy seed models this implies $z\sim5-12$, whereas for the light seed models this implies $z\sim8-12$.
\item $\overline{\lambda}$ is constant within a redshift bin and evolves according to $\overline{\lambda}=-2.49+1.93\log(1+z)$ \citep{Angles2015}, where $\lambda$ follows a log-normal distribution  as described in section 2.
\item We apply a minimum mass for a MBH according to the seeding mechanism: $5\times10^4\,$M$_{\odot}$ for heavy seeds, and $100$M$_{\odot}$ for light seeds.
\end{enumerate}
Figure \ref{fig:fxDist_torqueLimited} shows expected results for the heavy seed model X-ray flux distribution with observations from Lynx. A comparison to the heavy seed models presented in section 3 clearly demonstrates the ability to discriminate between the torque limited heavy seed model and other models presented in this work. In case of the torque limited light seed model, we do not expect to observe any MBHs at $z=8-12$ since it takes only $0.5\times t_{Hubble}$ for the MBHs to settle into the local scaling relation \citep{Angles2015}, and therefore this models are almost identical to the sub-Eddington $b=0$ light seed models  presented in section 3, for which no MBHs are expected to be observed at $z>10$ \citep[see also][]{Natarajan2017}.
\newpage
\section{Summary and Conclusions}
In this paper we investigated the expected X-ray flux distributions ($0.5-2\,$keV observed) from MBHs at high redshifts ($z>5$). We examined several models for SMBH birth and evolution: light seeds (Pop-III stars) and heavy seeds (direct collapse of a gas cloud) with sub-Eddington accretion, slim disk models allowing mild super-Eddington accretion (light seeds only), and torque limited accretion for both light and heavy seeds.
We assumed observations are taken with two experiments: the CDFS with a detection limit of $9.1\times10^{-18}\mathrm{erg\,cm^{-2}\,s^{-1}}$ at photon energies of $0.5-2\,$keV over a $5'\times5'$ FoV, and the proposed Lynx with a detection limit of $1\times10^{-19}\mathrm{erg\,cm^{-2}\,s^{-1}}$ at energies of $0.5-2\,$keV over a $400\,\mathrm{arcmin}^2$ FoV. For Chandra, such observations have been analyzed and so we can test our simulated results with respect to the existing analysis of the CDFS \citep[\textit{e.g.,}][]{Vito2013,Treister2013,Weigel2015}, while for deeper observations to be performed by future observatories our results are of a predictive nature. Our assumption is that the redshifts of observed objects will be known by complementary data at other wavebands guides us to consider only observations performed with instruments allowing high spatial resolution ($\sim1''$ or better), in order to avoid confusion with other targets in the beam \citep[such as the case with XMM-Newton and the Athena X-ray observatories, with $6''$ and $5''$ HPD respectively;][]{Arnaud2011,Collon2014}.

We assume that the observed MBHs are past the seeding stage and have commenced co-evolution with the host galaxy. As the MBHs and their hosts evolve towards the scaling relations observed in the local universe, a modified version of the scaling relation is realized at earlier times, which encodes physics of the seeding mechanism (\textit{e.g.,} Volonteri \& Natarajan 2009, Angles-Alcazr et al. 2013, 2015). After determining for each model the redshift dependent minimum $v_{circ}^{max}$ of a DM halo that hosts a detectable MBH, we used the Bolshoi N-body simulation to estimate the spatial density of detectable sources. We continued with a MC simulation for each model to account for the statistical distributions and uncertainties of the various parameters, and constructed the expected X-ray flux distributions to be observed with each X-ray telescope.

We found the number of high redshift MBHs with X-ray flux levels above the Chandra confusion limit is of order unity, statistically in agreement with zero observed sources. This is in agreement with current analysis of CDFS observations, as up to the time of the writing of this paper there is still no clear and unambigous detection of a high redshift MBH. These results clearly demonstrate the limits of current X-ray observatories. When assuming flux limits and FoV identical to the ones anticipated for the proposed Lynx, the number of expected observable MBHs rise considerably. More so, we find that the observed distributions from each family of models varies, and allows to limit the phase space in which the plausible scenario resides (or scenarios, in case more than one is realized in nature). We studied the expected contribution to the unresolved CXRB in each model, and found that in most of the simulated models the contribution lies below the upper limit set by CDFS measurements, even when considering the observed diminishing contribution from AGNs at low flux levels \citep{Hickox2006,Hickox2007,Lehmer2012}. Investigating the expected contribution to the CXRB from each model down to Lynx detection limit suggests that additional constraints are attainable by examining the flux level unaccounted for by resolved point sources.

Our simulated models make various assumptions about physical parameters affecting an MBH birth, evolution and emission - and while we use theoretically and observationally motivated arguments, the models are by no means exhaustive. As more data is accumulated about SMBHs in both the local and high redshift universe, and as a better theoretical understanding is obtained for various seeding and evolution models - one can further refine the underlying assumptions we make when simulating MBH flux and spatial distributions. Nevertheless, the results obtained in this work show the benefits, as well as the limitations, of X-ray observations of MBHs at high redshifts with existing and proposed experiments.
The recent discovery of gravitational waves from BH mergers have once again focused interest on the nature and evolution of BHs. While future gravitational wave observatories, such as LISA, expected to be launched at 2034\footnote{\textit{https://www.elisascience.org}}, will explore rapid events in the evolution of BHs (\textit{e.g.,} mergers), in this work we have explored what are the quasi-steady state observational signatures expected from MBHs as they evolve to the SMBHs found at the center of local galaxies. In that way, gravitational wave experiments and deep high resolution X-ray observations are complimentary approaches - both being crucial to our understanding of the evolution of the most massive objects known. The two approaches combined will allow us a better understanding of the physical processes responsible for the birth and evolutions of SMBHs which in turn will shed more light (figuratively speaking) on the emergence of structure in the universe less than a billion years after the big bang.\\

\end{document}